\documentclass[a4paper, 11pt]{article}
\usepackage[utf8]{inputenc}
\usepackage{jcappub} 
\usepackage{xcolor}
\usepackage[T1]{fontenc} 

\usepackage{amsmath, amssymb}
\usepackage{graphicx}
\usepackage{color}
\usepackage{sidecap}
\usepackage{bm}
\usepackage{multicol}
\usepackage{listings}
\usepackage{subcaption}
\usepackage[normalem]{ulem}
\usepackage{enumitem}

\setlist[itemize]{itemsep=1pt, topsep=3pt}
\let\ma\mathcal

\sidecaptionvpos{figure}{t}
\makeatother
\makeatletter
\newcommand*{\rom}[1]{\expandafter\@slowromancap\romannumeral #1@}
\makeatother


\title{\boldmath 
Cosmic strings, domain walls and environment-dependent clustering
}
\author[a]{\O{}yvind Christiansen,$^1$\note{Corresponding author.}}
\author[b,c,d]{Julian Adamek}
\author[d]{and Martin Kunz}

\affiliation[a]{CEICO, Institute of Physics of the Czech Academy of Sciences\\Na Slovance 1999/2, 182 00, Prague 8, Czechia}
\affiliation[b]{Institut f\"ur Astrophysik, Universit\"at Z\"urich\\ Winterthurerstrasse~190, 8057 Z\"urich, Switzerland}
\affiliation[c]{Institut f\"ur Teilchen- und Astrophysik, ETH Z\"urich\\ Wolfgang-Pauli-Strasse 27, 8093 Z\"urich, Switzerland}
\affiliation[d]{D\'epartement de Physique Th\'eorique, Universit\'e de Gen\`eve\\ 24 quai Ernest-Ansermet, 1211 Gen\`eve 4, Switzerland}

\emailAdd{oyvinch@fzu.cz}
\emailAdd{adamekj@ethz.ch}
\emailAdd{martin.kunz@unige.ch}
 
\abstract{
Recent cosmological data favour phantom-crossing dark energy, motivating models with non-minimal couplings that induce a fifth force on structure formation. Reconciling these models with local tests often requires strong screening, leading to environment-dependent clustering. We investigate such effects via a late-time structure-induced phase transition 
driven by a non-minimally coupled scalar field. For this purpose, we introduce \texttt{norns}, a fully relativistic cosmological particle-mesh code that self-consistently evolves a complex scalar field -- a generalisation of the symmetron producing global $U(1)$ strings rather than domain walls. Using  
simulations, we compare string and wall-forming models, quantifying impacts on the matter power spectrum, halo mass function, and defect dynamics. Strong environment-dependent effects can generate significant departures from $\Lambda$CDM in underdense regions while keeping the overall power spectrum changes modest ($\sim4$-$15\%$ at $k\sim0.3$-$0.5\,h\,\mathrm{Mpc}^{-1}$, sub-percent for $z \gtrsim 0.2$). We find that an attractive fifth force can locally suppress structure growth in voids while enhancing it in surrounding overdense regions by driving outflows from the voids. These effects leave distinctive signatures in the matter density probability density function and in marked halo power spectra, which are likely detectable in low-redshift data.
}

\begin{document}
\date{}
    \maketitle
    \flushbottom

    \section{Introduction}
    \label{S:introduction}

    The standard model of cosmology, $\Lambda$CDM, asserts that 95\% of the energy
    content of the late-time Universe resides in a \emph{dark} sector, consisting
    of a cosmological constant energy density, $\rho_{\Lambda}$, and cold dark matter
    (CDM). Their phenomenology is purely gravitational and explains the late-time
    accelerated expansion of space and observed features of gravitational
    structures and clustering. While the cosmological constant is especially poorly
    understood \citep{Zeldovich:1968ehl,Weinberg:1988cp}, progress
    on cold dark matter has historically been expected from particle accelerator
    experiments such as ATLAS and CMS, probing promising candidates within the
    framework of, for example, supersymmetry \citep{Martin:1997ns}.
    The lack of observations of effects from the dark sector on standard-model
    particles motivates a continued study into dark sector alternatives for
    model comparison and application to reported cosmological and astrophysical
    tensions.

    An interesting family of extended dark sector models introduces a phase
    transition in late-time cosmology. This has been shown to allow some improvement
    in joint analyses of cosmic microwave background (CMB) and large-scale structure
    (LSS) data \citep{Bassett:2002qu,Perivolaropoulos:2022txg,Alestas:2021nmi},
    and has also been studied in astrophysical contexts \citep{Burrage:2016yjm,Burrage:2018zuj,Kading:2023hdb}.
    A particular relativistic realisation of such a model is the symmetron \cite{Hinterbichler:2010es,Hinterbichler:2011ca},
    where an additional scalar field $\phi$ has a universal and non-minimal
    coupling to the stress energy of the matter sector. In the limit of small field
    excursions, the symmetron potential can be thought of as a Taylor polynomial
    and representative of a greater subset of the Horndeski theory landscape, encompassing
    models such as Brans-Dicke or dilatons which generically appear in the low-energy
    limit of string theory \citep{Tong:2009np}. The phase transition of the
    symmetron is controlled by the ambient stress energy of the matter fluid that
    maintains the symmetron's $\mathbb{Z}_{2}$ symmetry in overdense environments; therefore,
    the model can in principle comply with strict astrophysical constraints in the local neighbourhood
    \citep{Bertotti:2003rm,Esposito-Farese:2004azw,Tsujikawa:2008uc}.
    An indication for a \emph{phantom crossing} has recently been reported in
    the dark energy equation of state parameter
    \cite{DESI:2024hhd,DESI:2024aqx,Ishak:2024jhs}, for which,
    among others, non-minimally coupled models have been studied \citep{Chudaykin:2024gol,Ye:2024ywg,Wolf:2025jed},
    and the symmetron model in particular in \cite{Christiansen:2024hcc}. If the symmetron undergoes its phase transition at redshifts relevant for dynamical dark energy, around $z \sim 0.6$, then the small energy scale associated with the critical ambient density, $\rho_*$, implies that the field must undergo large field excursions to reach an energy scale comparable to dark energy. This in turn produces very strong fifth forces, which must be countered by correspondingly strong screening effects. The Structure-Induced Phase Transition (SIPT), found in \cite{Christiansen:2024vqv} and discussed below, is especially interesting in this context. SIPT-type models may be applied to address the coincidence problem of dark energy, by relating a late-time phase transition to nonlinear structure formation. As such, it can be seen as an extension on back-reaction models, that have been shown to not work in standard $\Lambda$CDM (e.g. \cite{Adamek:2018rru}). As we will see in Section \ref{S:method}, reaching energy-scales relevant to dark energy in the SIPT-regime of the parameter space requires a generalisation of the symmetron model, by e.g. introducing non-perturbative conformal factors or potentials. 

    The symmetron adds a single additional scalar field, $\phi\in \mathbb{R}^{1}$,
    where a spontaneously broken $\mathbb{Z}_{2}$ symmetry results in the
    formation of domain walls between two degenerate vacua. By choosing the field
    to be complex, $\phi\in \mathbb{C}^{1}$, it breaks a global $U(1)$
    symmetry, resulting in string defects forming at the locations of phase discontinuities.
    The respective defects can be formed by the Kibble mechanism \citep{Vilenkin:1984ib}
    or as a result of phase transitions occurring independently in different underdense
    patches of the Universe – the case of SIPT. One can similarly produce monopoles or textures by
    promoting the scalar field to a 3- or 4-vector, respectively. For more
    information on the application of topological defects in cosmology, we direct
    the reader to the useful reviews \cite{Hindmarsh:1994re,Vilenkin:1984ib,Vachaspati:1984dz,Durrer:2001cg,Vachaspati:2006zz}.
    Recently, in \cite{Nezhadsafavi:2025pzg}, the complex symmetron was
    considered in a simple ideal scenario to study the ability of overdensities
    to pin and stabilise strings, as has been shown for the case of domain walls
    in \cite{Llinares:2014zxa}.

    Here, for the first time, we perform cosmological simulations of non-minimally
    coupled cosmic strings in the late-time Universe, and show side by side
    comparisons to the domain-wall scenario of the real-valued symmetron field. We
    demonstrate differences in their effects on matter clustering, dark-matter
    halo statistics, defect dynamics, and their pinning to overdensities. We
    provide estimates of fifth-force effects on dark-matter halos, particle and halo
    velocity dispersions, and discuss target observables for constraining the model
    in upcoming experiments. In particular, we demonstrate strong signals in the
    matter density probability function and in marked statistics, and discuss them in relation to ongoing galaxy surveys. We provide
    visualisations of the simulation data in interactive figures and animations.\footnote{\url{https://www.oyvindchristiansen.com/projects/rcsymmetron
    }
    \label{footnote:animations} } 

    The simulation code was developed based on the scheme of \texttt{ISIS} \cite{Llinares:2013qbh,Llinares:2013jza,Llinares:2013jua}
    in which the real-valued symmetron is simulated on a lattice on top of the Newtonian
    N-body code \texttt{RAMSES} \citep{Teyssier:2001cp}. In \texttt{asevolution}
    \cite{Christiansen:2023tfy,Christiansen:2024vqv}, this
    method was implemented in the fully relativistic weak-field particle-mesh
    code \texttt{gevolution} \cite{Adamek:2015eda,Adamek:2016zes,Adamek:2014xba,Daverio:2015ryl},
    which allows to consistently treat dark sector energy transfer, back-reaction
    effects and relativistic particle velocities. It was additionally extended to evolve
    gravitational waves in \texttt{AsGRD} \citep{Christiansen:2024uyr}.
    Large scalar masses $\mu\gg \left(D/1\,\rm{ m}\right) \times 0.1$ neV, can
    be constrained in laboratory experiments \citep{Llinares:2018mzl},
    where $D$ is the scale of the experimental setup, while cosmologically relevant
    Compton wavelengths, $L_{\mathrm{C}}$, are constrained by solar system and galactic
    observations \citep{Hinterbichler:2010es}. In \cite{Christiansen:2024vqv},
    the SIPT-regime of the symmetron parameter space was explored that allows for
    inhomogeneous phase transitions within underdense voids. This part of the parameter
    space presents interesting and novel phenomenology, is relatively poorly
    constrained as the usual modelling does not apply -- see e.g.\ figure 5 of
    \cite{Burrage:2023eol} for $\rho_{\rm{SSB}}<\bar\rho$ -- and will be the
    target for this work as well.

    The simulation code \texttt{norns} (NOn-linear and Relativistic Numerical Strings)  is made publicly available,\footnote{
    symmetron is done with \texttt{AsGRD}.\footnote{\url{https://github.com/oyvach/AsGRD}\label{footnote:AsGRD_code}} Convergence and resolution tests, as well as code performance are discussed in appendices~\ref{A:stringProfile}-\ref{A:performance}.
    In addition to evolving the symmetron, the simulation codes present a simple
    and rigorous framework for the inclusion of physics beyond the standard
    model into nonlinear structure formation. In the past, \texttt{gevolution} has
    also been applied, for example, to $k$-essence and KGB dark energy and $f(R)$
    gravity \citep{Hassani:2019lmy,Nouri-Zonoz:2025cul,Reverberi:2019bov}.

    \section{Method}
    \label{S:method}

    The implementation scheme generally matches that presented in \cite{Christiansen:2023tfy,Christiansen:2024vqv},
    but with the exception of the promotion of the symmetron to complex-valued
    $\phi\in\mathbb{C}$. We give a short overview of the new methodological
 aspects here, while the reader is directed to the references for the remainder of
    the treatment. We detail the post-processing methodology for the statistics that
    we present together with the results in Section \ref{S:results}.

    \subsection{Model}
    \label{SS:model}

    The symmetron model is formulated in the Einstein frame with a Klein-Gordon
    plus a self-interaction potential. The metrics of the Einstein and Jordan frames,
    $g_{\mu\nu}$ and $\tilde{g}_{\mu\nu}$, respectively, are related by a
    conformal transformation as $\tilde g_{\mu\nu}=A^{2}(\phi) g_{\mu\nu}$. In the
    Jordan frame, the scalar is uncoupled to the matter sector but achieves a
    universal matter coupling on transformation to the Einstein frame. The matter
    Lagrangian is written in the Jordan frame and is $\ma L_{\mathrm{m}}(\tilde{g}
    _{\mu\nu},\Psi_{i})$, for the standard model fields $\Psi_{i}$. On varying with
    respect to $\phi$, it generates a term $-\frac{A_{\phi}}{A}T_{\mathrm{m}}$, where
    $T_{\mathrm{m}}$ is the trace of the stress energy of the Einstein frame matter
    sector and is equal to the negative of the matter energy density $T_{\mathrm{m}}
    \approx -\rho_{\mathrm{m}}$ in the nonrelativistic limit. This allows us to
    write an effective potential from which the equations of motion are easily
    obtained,
    \begin{align}
        V_{\rm{eff}}(\phi) = -\frac{1}{2}\mu^{2} \phi^{\dagger} \phi + \frac{\lambda}{4}\left(\phi^{\dagger}\phi\right)^{2} - T_{\mathrm{m}}\ln \left(A[\phi]\right).
    \end{align}
    We consider both cases $\phi\in\mathbb{C},\mathbb{R}$. Now, the full effective
    Lagrangian density for the symmetron is given by
    \begin{align}
        \mathcal{L}= - \frac{1}{2}\left(\partial_{\mu} \phi\right)^{\dagger} \left(\partial^{\mu}\phi\right) - V_{\rm{eff}}(\phi).
    \end{align}
    We also define
    \begin{align}
        V_{0}   & \equiv \frac{\lambda}{4}v_{0}^{4} = \frac{9 \Omega_{\mathrm{m},0}^{2} H_{0}^{2}}{2 a_{*}^{3}}\frac{\beta^{2}\xi_{*}^{2}}{a_{*}^{3}}= \frac{9 \Omega_{\mathrm{m},0}^{2} H_{0}^{2}}{2 a_{*}^{3}}\Delta A_{\mathrm{max}}, \label{eq:V0} \\
        \rm d A & \equiv A-1 = \frac{1}{2}\left( \frac{\phi}{M}\right)^{2} \lesssim \frac{\beta^{2} \xi_{*}^{2}}{a_{*}^{3}}, \label{eq:dA}
    \end{align}
    where $V_{0}$ is the energy difference between the false vacuum ($\phi=0$) and
    the true vacua ($\phi=\pm v_{0}$), and we have Taylor expanded the potential
    $V(\phi)$ and the conformal factor $A(\phi)$, assuming $\mathbb{Z}_{2}$ symmetry
    on $\phi\rightarrow -\phi$, and small field excursions
    $\rm d A,|\phi/M_{\rm{pl}}|\ll 1$, where $M_{\rm{pl}}$ is the Planck mass. $\Delta A_{\mathrm{max}}=\beta^2\xi^2_*/a_*^3$ is the deviation of the conformal factor from unity in the true vacuum (where $T_\mathrm m=0$).

    The model has three free parameters: the bare tachyonic mass $\mu$, the self-interaction
    strength $\lambda$ and the conformal coupling $M$. These are analogous to
    the phenomenologically related quantities of the Compton wavelength compared
    to the current time Hubble length $\xi_{*}\equiv L_{\mathrm{C}}\,H_{0}$, the scale factor
    of the phase transition in a homogeneous universe $a_{*}=1/(1+z_{*})$ and the
    fifth force strength relative to the Newtonian gravitational force for small
    perturbations around the true vacuum $F_{5}/F_{N}\simeq 2\beta^{2}$. They
    are explained in more detail in
    \cite{Christiansen:2023tfy,christiansen_cosmological_2024-1}.

    From Eq.~\eqref{eq:V0}, we observe that if we require $\Delta A\lesssim 0.1$
    to remain in the perturbative regime, which currently is a limitation of our weak-field
    expanded scheme \citep{Christiansen:2023tfy,Adamek:2016zes},
    we have
    \begin{align}
        \frac{V_{0}}{\rho_{c,0}}\sim 1.5\times 10^{-2} \left(\frac{\Delta A}{0.1}\right)  a_{*}^{-3}.
    \end{align}
    This means that we cannot have a significant effect on the background in the
    perturbative $\Delta A$ regime for the interesting case of SIPT\footnote{We call the phase transition `structure-induced' when $z_{\rm{SSB}}\gg z_{*}$, where $z_{\rm{SSB}}$ is the actually
    observed redshift of the phase transition. In a homogeneous universe $z_{\rm{SSB}}=z_*$.} that is also accessible
    to our simulations. The background evolution therefore remains well
    approximated by $\Lambda$CDM in this case. However, we note that the Universe
    is significantly more inhomogeneous at early redshifts for smaller smoothing
    scales $L_{\mathrm{C}}\ll 1\,\mathrm{Mpc}$, which is expected to allow for
    the SIPT type of phase transition at smaller $a_{*}$. Some of the
    phenomenology explored here might also be taken as representative for this regime.
    Treating non-perturbative $\Delta A$ is beyond the scope of the current work
    and requires a generalisation of the particle geodesic solver \citep{Christiansen:2023tfy}.
    For steeper barriers, for example, when the bounding potential term contains
    $\phi^{N},\, N>4$, one can have a larger $V_{0}$ while keeping $v_{0}$ and thus
    $\rm d A(v_{0})$ (also $\beta\sim v/M^{2})$ small. With a sine-Gordon potential
    in particular $V\supset \mu^{2}(1-\cos(\phi/f))/2$, the minimum and
    amplitude can be set separately, and therefore allow large energy contents in
    the symmetron sector while keeping conformal factors perturbative and fifth forces
    either large or small. Such potentials have particular relevance for axion
    dark matter models \cite{Choi:2020rgn}, and will be the subject of a future
    project. For now, we keep the symmetron potential at quartic order, and use it as a proxy for studying environment-dependent signatures representing a broader class of SIPT-models with different potentials, conformal factors, kinetic structure, non-universal couplings or multiple fields. 

    \subsection{Defect finding}
    \label{SS:defectFinding}

    In order to identify strings in the simulations, we follow the procedure laid
    out in \cite{Vachaspati:1984dz}, see Fig.~\ref{fig:string_schematic}.
    If a string passes through a region, then the closed line integral along a path
    enclosing the string will give the string's `winding number' $N$
    \begin{align}
        N = \frac{1}{2\pi}\oint \rm d \varphi ,
    \end{align}
    where $\varphi$ is the complex phase. Therefore,
    an indication that the region enclosed contains a string is found by finding
    the shortest distance around the unit circle of pairwise points, as $|\rm{diff}(\theta_1,\theta_2)|=|\theta_{1   }-\theta
    _{2}|\leq \pi$. The phase increments are then summed up along a closed path of $4$ neighbouring vertices, see Fig.~\ref{fig:string_schematic}. We define counter-clockwise rotations as positive. Algorithmically
    (indicating translations along axes $x,y,z$ from a point $\vec p$ by $\vec p_x$, $\vec p_y$, $\vec p_z$
    respectively) 
\begin{lstlisting}
for (*@$\vec p$@*) in points
    (*@$\mathrm{path}_{yz} = \mathrm{diff}(\vec p,\vec p_{y}) + \mathrm{diff}(\vec p_{y},\vec p_{yz}) + \mathrm{diff}(\vec p_{yz},\vec p_{z}) + \mathrm{diff}(\vec p_{z},\vec p)$@*)
    (*@$\mathrm{path}_{zx} = \mathrm{diff}(\vec p,\vec p_{z}) + \mathrm{diff}(\vec p_{z},\vec p_{zx}) + \mathrm{diff}(\vec p_{zx},\vec p_{x}) + \mathrm{diff}(\vec p_{x},\vec p)$@*)
    (*@$\mathrm{path}_{xy} = \mathrm{diff}(\vec p,\vec p_{x}) + \mathrm{diff}(\vec p_{x},\vec p_{xy}) + \mathrm{diff}(\vec p_{xy},\vec p_{y}) + \mathrm{diff}(\vec p_{y},\vec p)$@*)
    (*@${\text{if path}}_{xy} > \pi$@*)
        (*@$\mathrm{string}_{xy}(\vec p) = 1$@*)
    (*@${\text{else if path}}_{xy} < -\pi$@*)
        (*@$\mathrm{string}_{xy}(\vec p) = -1$@*)
    ...
\end{lstlisting}
    \begin{figure}[tb]
        \centering
        \includegraphics[width=0.4\linewidth]{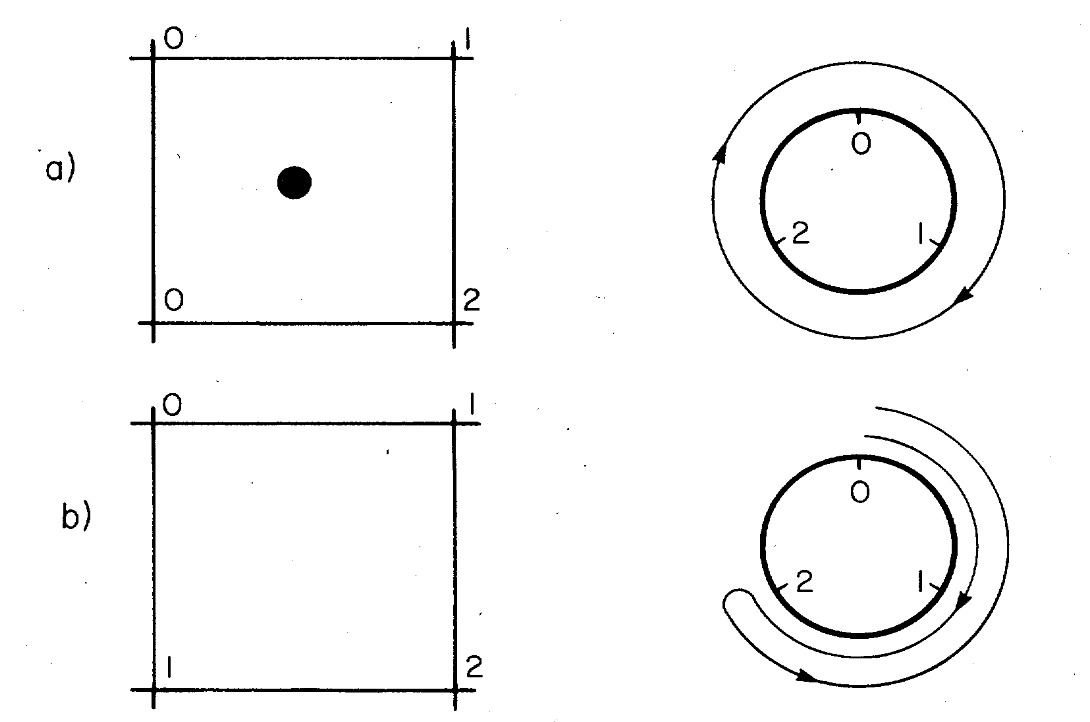}
        \caption{Schematic from \cite{Vachaspati:1984dz}, indicating the
        complex phase at four different vertices enclosing a) a string, or b)
        nothing. Right panel indicates how the path integral goes in either case.
        When the phase varies continuously there are no phase defects.}
        \label{fig:string_schematic}
    \end{figure}
    Using 4 points only, this algorithm is unsuitable for resolving winding numbers $N>1$. Larger winding numbers are however expected to be rare and short-lived \cite{Hindmarsh:1994re}, and therefore not important to keep track of here.
    A similar scheme is applied for finding domain walls in the $\phi\in\mathbb{R}$
    simulations, but instead of looking for phase discontinuities, the locations, $\vec p$,
    where the field $\chi$ crosses $0$ are identified by the criterion $\mathrm{sign}
    (\chi(\vec p_{-i})\chi(\vec p_{i}))=\mathrm{sign}(\chi(\vec p)\chi(\vec p_{i}))<0$, where
    $i\in\{x,y,z\}$ indicates which axis we are comparing along, and $\pm i$ indicates positive or negative increments along this axis. We make sure that it is $\chi(\vec p)\chi(\vec p_{i})$ that has
    different sign among the three in order not to double count the walls at neighbouring
    points. By comparing three adjacent vertices in the grid, the wall finder algorithm
    is more robust against spurious detections; in the deeply screened limits or before the phase transition
    where the field is very small, we expect
    numerical errors to give some isolated spurious defect detections in both cases of
    a real and a complex field.

    \subsection{Parameter choices}
    \label{SS:parameters}

    As mentioned in Section \ref{S:introduction}, for the current
    work, we are interested in the regime of the structure-induced phase transition
    ($z_{\rm{SSB}}\gg z_{*}$) reported in \cite{Christiansen:2024vqv}. In particular,
    we target the parts of the parameter space where $z_{\rm{SSB}}$, being the redshift
    of the observed phase transition, is $0.5 \lesssim z_{\rm{SSB}}\lesssim 0.8$.
    We can have a rough prediction for $z_{\rm{SSB}}$ by running smaller box size
    simulations at the same resolution of $\mathrm{d x}\sim 0.4\,h^{-1}\,\mathrm{Mpc}$.
    However, the range of underdensities at the smoothing scale of $L_{\mathrm{C}}$
    will be increasingly limited by sample variance, which will leave some
    uncertainty for the actual $z_{\rm{SSB}}$ in the larger box simulation. This
    can be alleviated somewhat by running multiple exploratory small box simulations
    with different seed numbers.

    The structure-induced, inhomogeneous phase transition, where $z_{\mathrm{SSB}}
    \neq z_{*}$, occurs when there are significant underdensities at scales
    $\gg L_{\mathrm{C}}$. In practice, for
    $L_{\mathrm{C}}\sim 1\,h^{-1}\,\mathrm{Mpc}$, this happens for choices $z_{*}
    \lesssim 1$. The smaller $z_{*}$, the stronger the screening in overdensities.
    Being interested in strong screening and environment-dependent effects, we put $z_{*}
    \leq 0.1$. This becomes increasingly computationally expensive as smaller
    $z_{*}$ means larger effective masses in overdensities, which requires resolving
    smaller timesteps to capture the phase evolution of the field. Meanwhile, for
    smaller $z_{*}$, one also needs smaller $L_{\mathrm{C}}$ to undergo phase
    transition at a similar time, which sets a larger mass for the field in underdensities,
    requiring smaller timesteps to resolve the phase evolution there as well, in
    addition to requiring better spatial resolution. See
    \cite{Christiansen:2024vqv} for more details. We are limited to $L_{\mathrm{C}}
    > \rm d x$ and a time step for the scalar field $\rm d \tau \ll \rm d x$ to
    ensure the stability of the time integration.

    The final parameter that we fix is $\beta$, which only impacts the vacuum-normalised
    scalar field's evolution indirectly through the change in the matter
    configuration; it increases the enhanced clustering of the matter field in the
    unscreened regions and pushes more matter into the screened regions. As our
    aim is a proof-of-concept exploration of this lesser known part of the
    symmetron parameter space, we make sure to pick $\beta$ large enough to have
    significant $\sim 4$-$15\%$ effects on the matter power spectrum. In
    practice, it is difficult to establish the exact effect on the matter power spectrum
    due to the abovementioned sample variance in the smaller tentative box simulations.
    As we shall see later, simulation $5_{\mathbb{C}}$ has a very large effect $\Delta
    P/P\sim 1$, but may still provide phenomenologically interesting
    observations about the symmetron, relevant to more viable parameter choices.

    The final parameter choices are shown in Table~\ref{tab:parameters}, where
    we provide both the choice of phenomenological parameters ($z_{*}$, $\beta$,
    $L_{\mathrm{C}}$) and the equivalent choice of Lagrangian parameters ($\mu$,
    $M$, $\lambda$) in the units of electronvolts (eV). We also show the difference
    of the conformal factor from unity, $\rm d A$, Eq.~\eqref{eq:dA}, and the
    size of the symmetron energy barrier, $V_{0}$, Eq.~\eqref{eq:V0}, setting the
    overall energy scale of the symmetron sector. The remaining cosmological parameters are chosen according to the Planck $\Lambda$CDM best-fit parameters \cite{Planck:2018vyg}, but with number of relativistic species $N_{\mathrm{NCDM}}=0$ and $N_{\mathrm{eff}}=3.046$. All simulations parameters will be made available on zenodo after peer-review.

    \begin{table}[t]
        \centering
        \begin{tabular}{|c|c|c|c|c|c|c|c|c|c|c}
            \hline
            model            & $z_{*}$ & $\beta$ & $L_{\mathrm{C}}$ {\tiny[$h^{-1}\,\mathrm{Mpc}$]} & $\mu$ {\tiny{[$10^{-30}\,$eV]}} & $M$ {\tiny{[$10^{24}\,$eV]}} & $\lambda$\tiny{$\,\cdot 10^{105}$ } & $\rm d A_{\phi=v_0}$\tiny{$\,\cdot 10^{5}$} & $V_{0}$\tiny{$\,\cdot 10^{12}/\rho_{c,0}$} \\
            \hline
            1$_{\mathbb{C}}$ & $0.1$     & 16          & 1                                                & 3.16                            & 6.3                          & 3.7                                 & $1.9$                                       & $2.8$                                      \\
            $1_{\mathbb{R}}$ & $0.1$     & 16          & 1                                                & 3.16                            & 6.3                          & 3.7                                 & $1.9$                                       & $2.8$                                      \\
            2$_{\mathbb{C}}$ & $0$       & 50          & 0.66                                             & 4.8                             & 3.6                          & 8.1                                 & $6.1$                                       & $6.7$                                      \\
            $2_{\mathbb{R}}$ & $0$       & 50          & 0.66                                             & 4.8                             & 3.6                          & 8.1                                 & $6.1$                                       & $6.7$                                      \\
            3$_{\mathbb{R}}$ & $-0.1$    & 50          & 0.45                                             & 7                               & 2.1                          & 150                                 & $2.1$                                       & $1.7$                                      \\
            4$_{\mathbb{C}}$ & $-0.1$    & 100         & 0.45                                             & 7                               & 2.1                          & 38                                  & $8.2$                                       & $6.7$                                      \\
            5$_{\mathbb{C}}$ & $-1/3$    & 1000        & 0.45                                             & 7                               & 1.3                          & 2.3                                 & $410$                                       & $170$                                      \\
            \hline
        \end{tabular}
        \caption{Choices of parameters for the suite of simulations, using boxsize
        $B=500\,h^{-1}\,\mathrm{Mpc}$, and number of grids/particles
        $N=1280^{3}$. The sets of parameters $z_{*},\,\beta,\,L_{\mathrm{C}}$
        and $\mu,\,M,\,\lambda$ can be mapped onto one another one-to-one, and
        are both stated here for convenience.}
        \label{tab:parameters}
    \end{table}
\newpage
    \section{Results}
    \label{S:results}

    \paragraph{Overview:}
    We show the evolution of the fraction of the simulation volume that has phase transitioned
    in Fig.~\ref{fig:overview}. The general response to variation of the
    phenomenological parameters ($L_{\mathrm{C}},z_{*},\beta$) is similar to
    that in the real-valued symmetron case, as shown in \citep{Christiansen:2023tfy}.
    \begin{figure}[tb]
        \centering
        \includegraphics[width=0.7\linewidth]{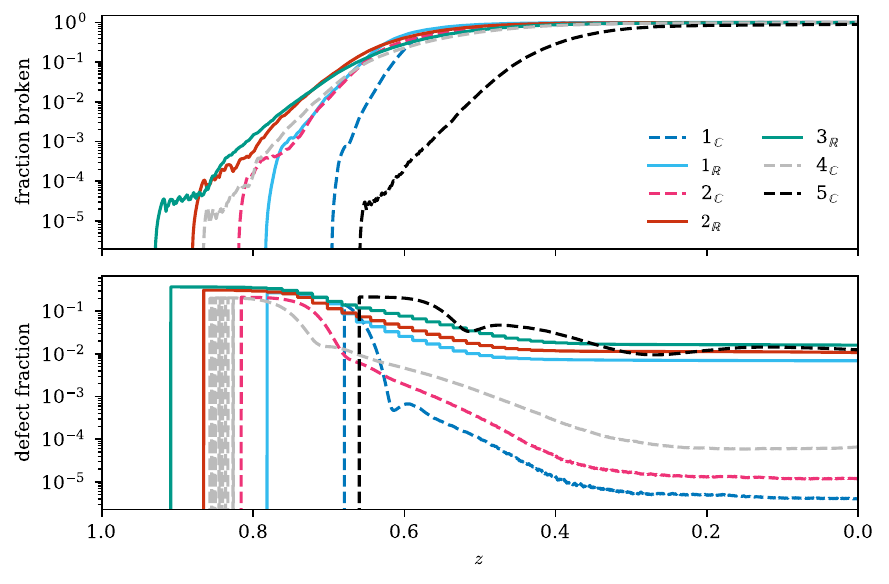}
        \caption{ Top: Fraction of the simulation volume where the global
        $U(1)$ or $\mathbb{Z}_{2}$ symmetry is broken, approximated by $|\Re
        (\chi)|>10^{-2}$. Bottom: Fraction of simulation volume where phase discontinuities
        or sign switches are found, respectively in the case of
        $\phi \in \mathbb{C}$ or $\phi \in \mathbb{R}$.}
        \label{fig:overview}
    \end{figure}

    \begin{figure}[t]
        \centering
        \begin{subfigure}
            [b]{0.48\textwidth}
            \centering
            \includegraphics[width=\textwidth]{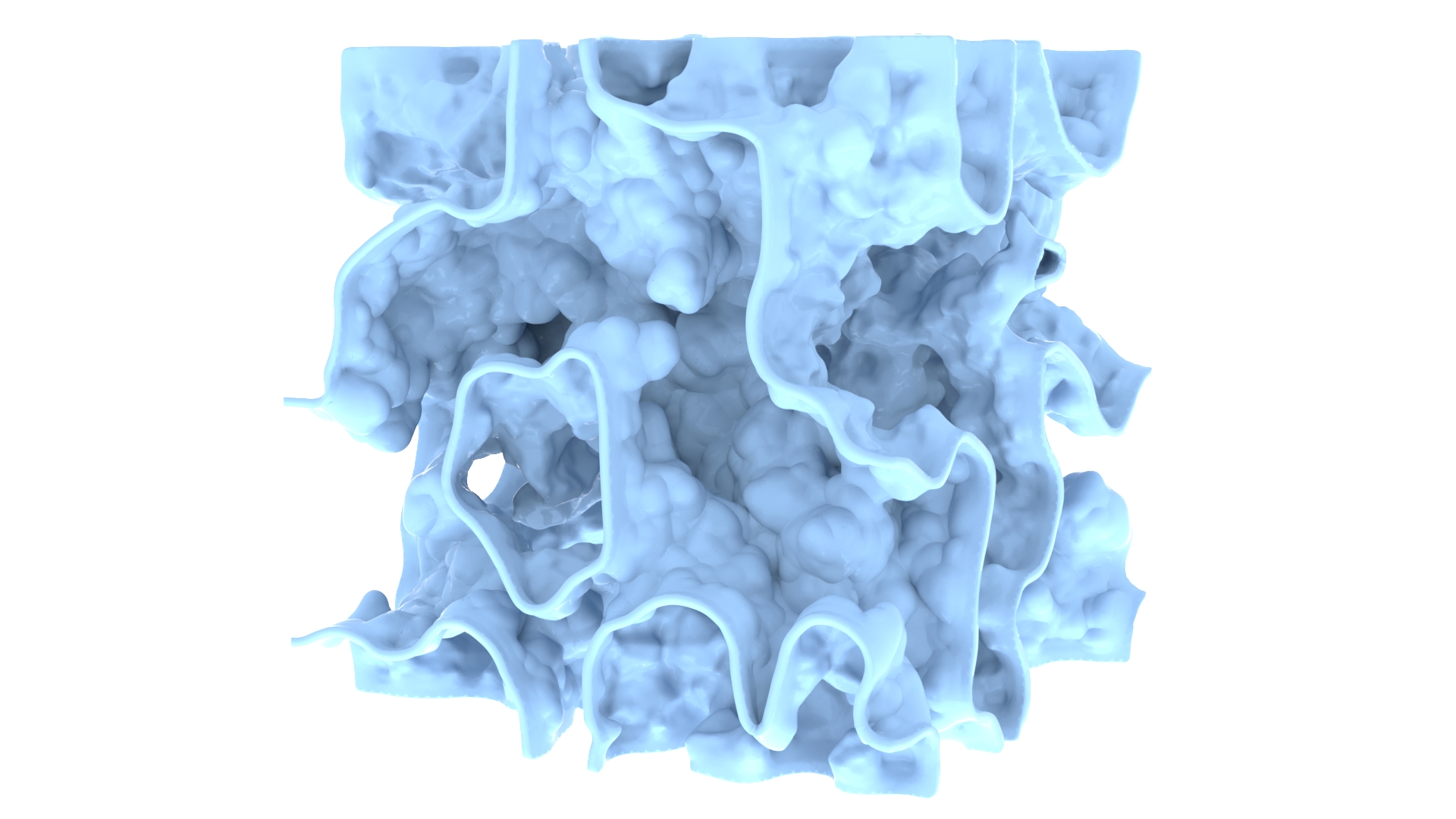}
        \end{subfigure}
        \hspace{-0.1\textwidth}
        \begin{subfigure}
            [b]{0.48\textwidth}
            \centering
            \includegraphics[width=\textwidth]{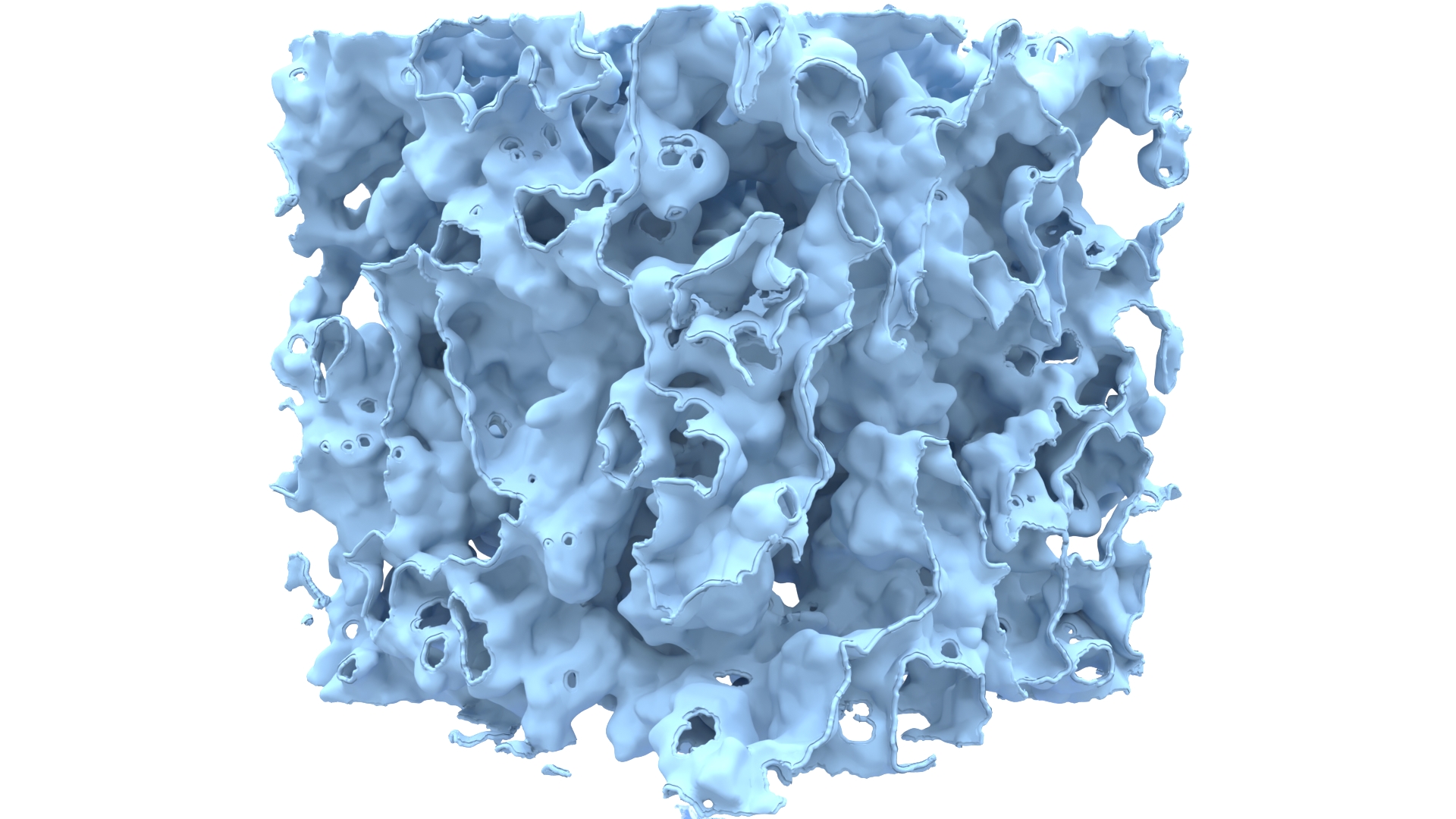}
        \end{subfigure}
        \\
        \vspace{0.01\textwidth}
        \begin{subfigure}
            [b]{0.48\textwidth}
            \centering
            \includegraphics[width=\textwidth]{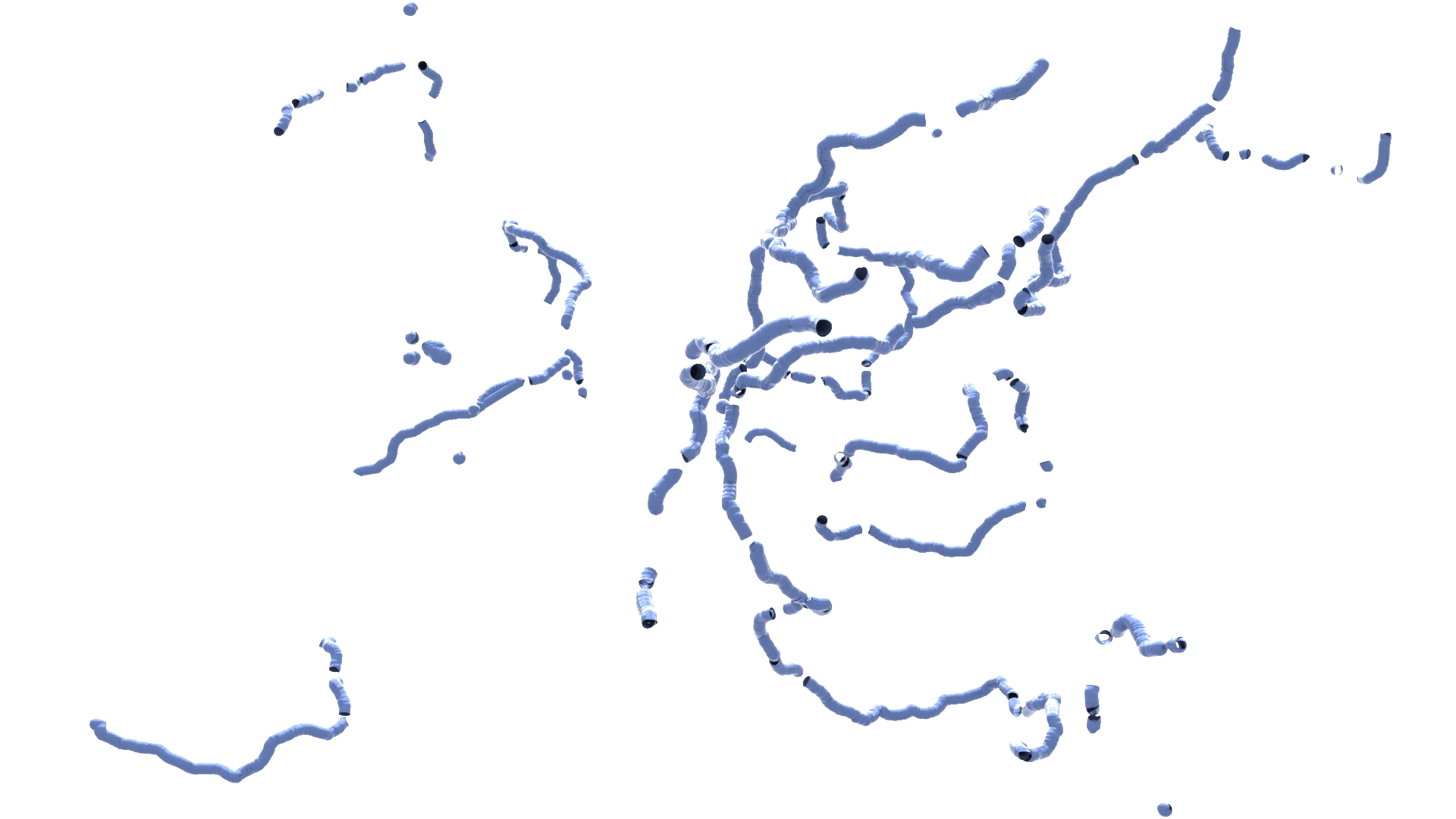}
        \end{subfigure}
        \hspace{-0.1\textwidth}
        \vspace{0.03\textwidth}
        \begin{subfigure}
            [b]{0.48\textwidth}
            \centering
            \includegraphics[width=\textwidth]{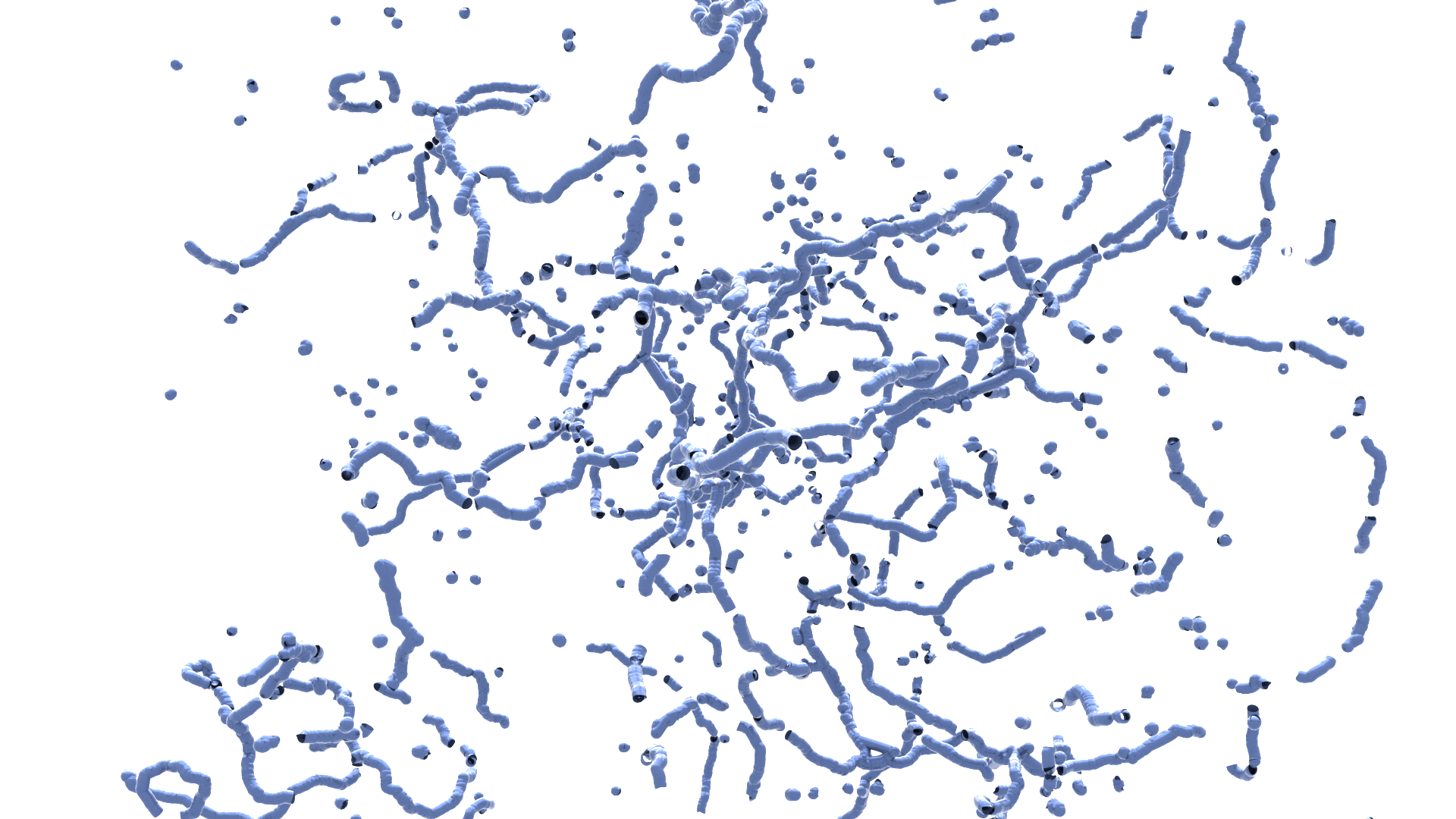}
        \end{subfigure}
        \caption{The defect structures that were identified at redshift $z=0$
        for simulations (from top left to bottom right) $1_{\mathbb{R}}$, $2_{\mathbb{R}}$,
        $1_{\mathbb{C}}$, and $2_{\mathbb{C}}$. }
        \label{fig:defect_renderings}
    \end{figure}

    \paragraph{Defect evolution:}
    The evolution of the defect density of the symmetron is shown here for the
    first time, in Fig.~\ref{fig:overview}. In general, the phase field is
    smooth before the phase transition, but the defect finder finds a very large defect
    density in the box as the phase transition occurs\footnote{The initial
    plateau is converged in spatial resolution if assuming that initial defects are 3-dimensional objects, that then evolve into 1-dimensional structures; see appendix \ref{A:convergence}.}. This is followed by a rapid decrease,
    roughly coinciding with the start of rapid growth of the volume of broken
    phase (indicated by the field being larger than a threshold value $|\phi|\gtrsim
    10^{-2}$). This continues until some lower density plateau, where the
    defects have taken on a large scale structure as seen in Fig.~\ref{fig:defect_renderings},
    which occupies more cells in the cases of domain walls. In the very early phase
    transition, the defect finder finds mostly isolated and disconnected
    segments of points that fill the volume. They are slowly sorted out by self-annihilation
    or mergers, leaving in the end only a fraction of more topologically stable
    extended strings. The process is similar for the case of domain walls. For
    the strong screening cases of models 3-5, this transition is slowed down,
    and the defects are still somewhat unformed at redshift $z=0$, see Fig.~\ref{fig:stringsEvolutionB}.
    This is assisted by the stronger pinning to overdensities that keeps the strings
    from moving. As time goes on, the defects both move into and become pinned to
    overdensities, and enhance clustering at their locations. Defects that remain
    in underdensities without pinning are also unstable and more likely to annihilate.
    We therefore see in Fig.~\ref{fig:stringenvhistogram} that the probabilities
    of the density at the locations of defects start out dominated by underdensities
    but then move towards larger densities with time. The unnormalised
    histograms show that this is both because of a rapid decline of defects at
    underdensities and a steady growth of number of defects at overdensities, consistent
    with what Fig.~\ref{fig:stringsEvolutionB} is showing.

    \begin{figure}[t]
        \centering
        \includegraphics[width=\linewidth]{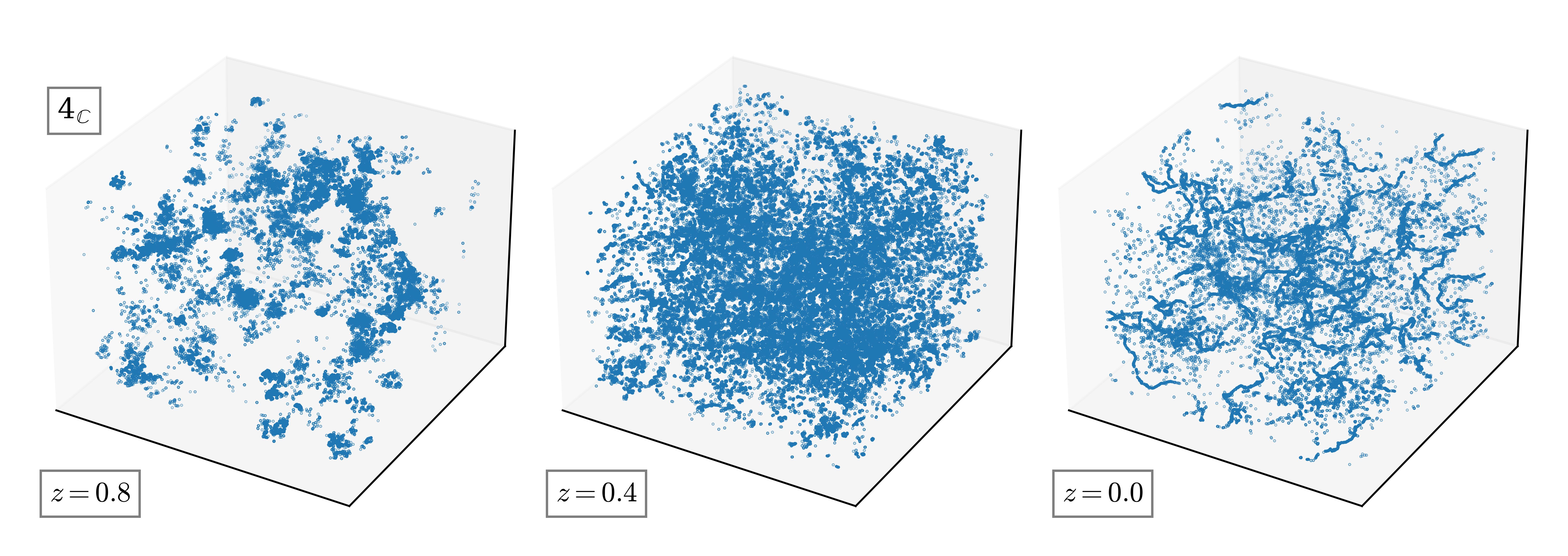}
        \caption{String defects in model $4_{\mathbb{C}}$ at three different
        redshifts. Strong screening/pinning increases the time for the strings to
        resolve themselves and assume global structure.}
        \label{fig:stringsEvolutionB}
    \end{figure}
    \begin{figure}[t]
        \centering
        \includegraphics[width=\linewidth]{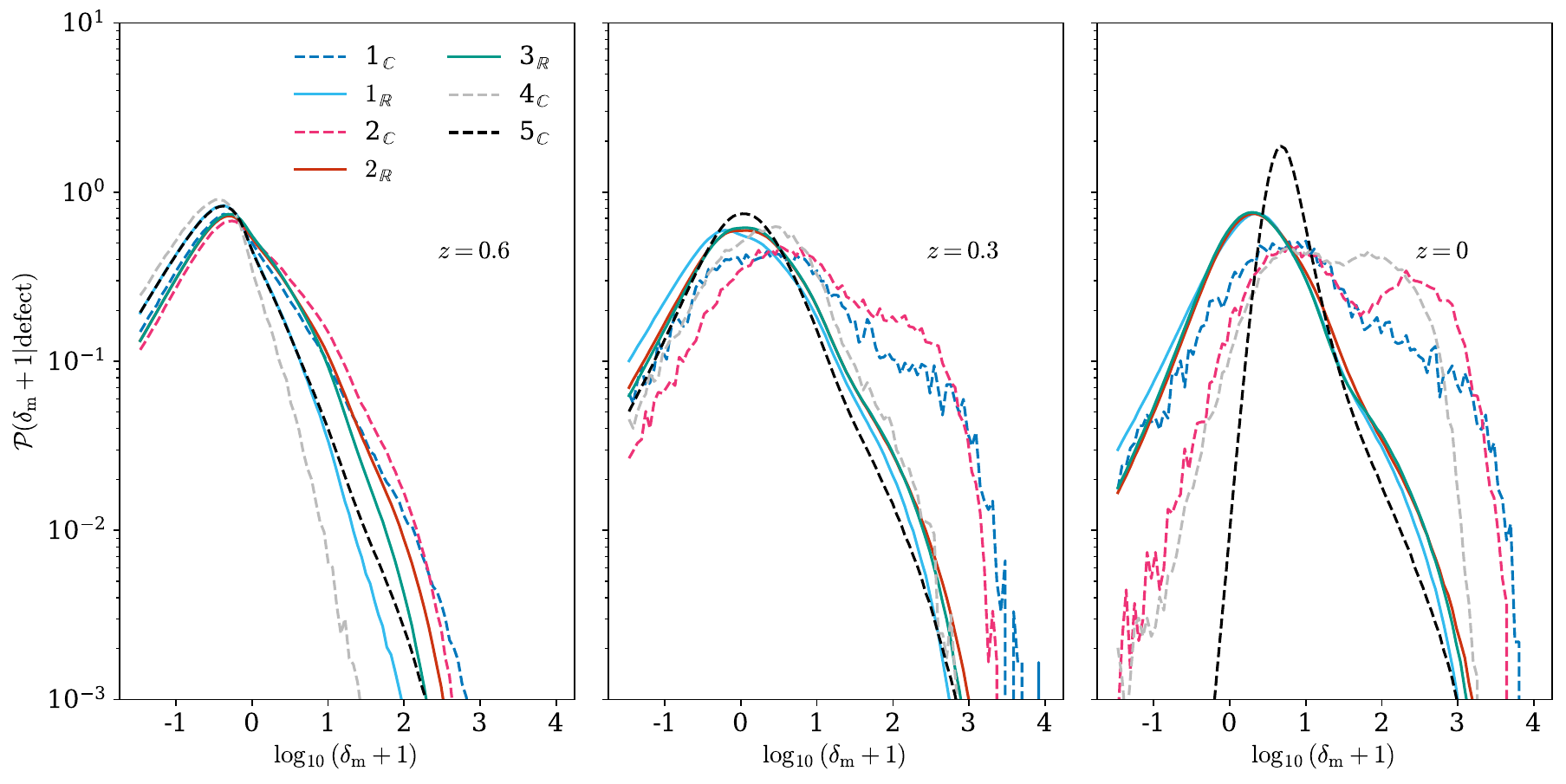}
        \caption{Histograms of the matter overdensity $\delta_{m}$, at the locations
        of the identified defects, for different simulations and redshifts. }
        \label{fig:stringenvhistogram}
    \end{figure}

    \begin{figure}[t]
        \centering
        \includegraphics[width=\linewidth]{
            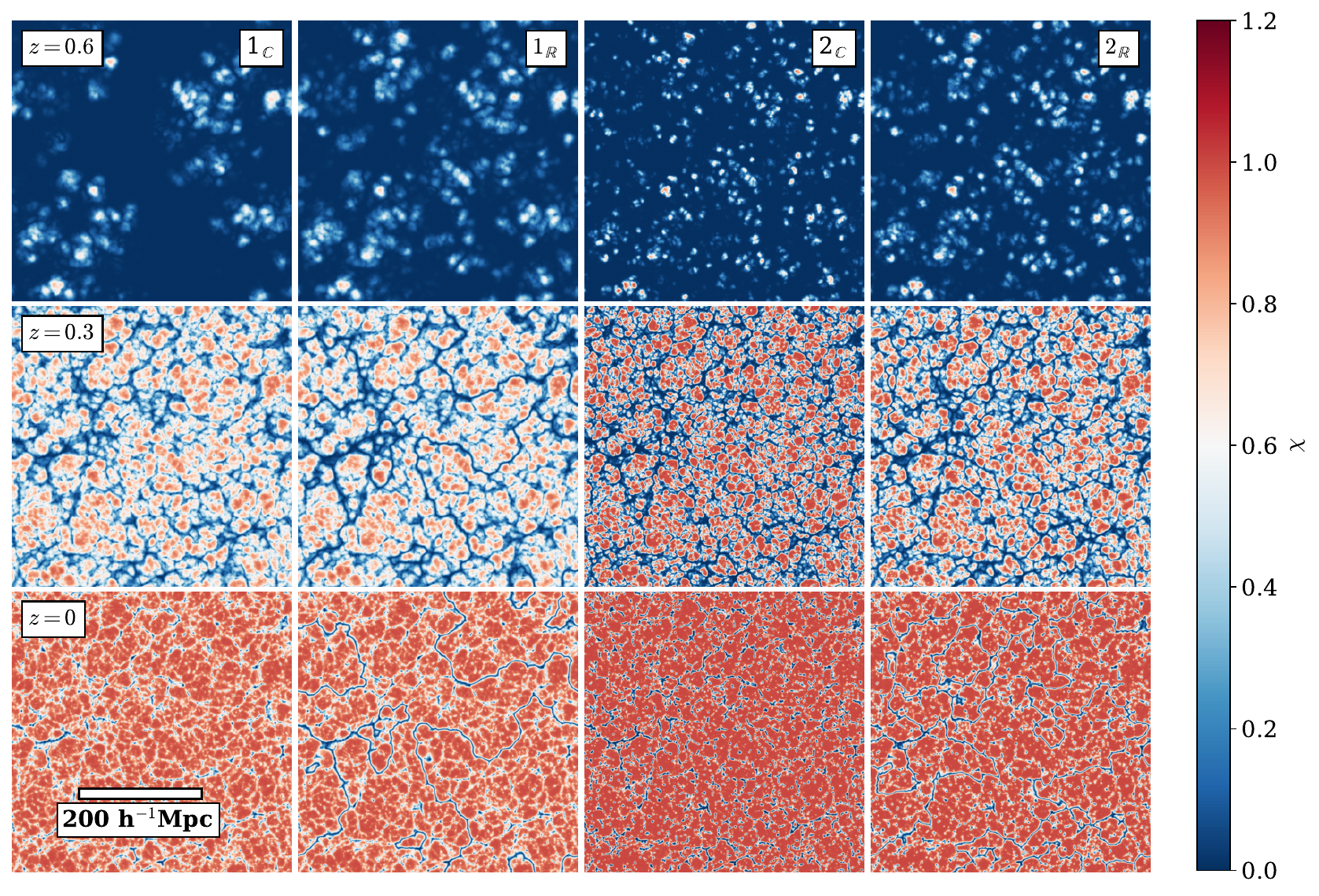
        }
        \caption{The norm of the scalar field $|\chi|$ for different simulations
        and redshifts, shown on a slice of the simulation volume.}
        \label{fig:fieldcartoon1-2}
    \end{figure}
    \begin{figure}[ht!]
        \centering
        \includegraphics[width=\linewidth]{
            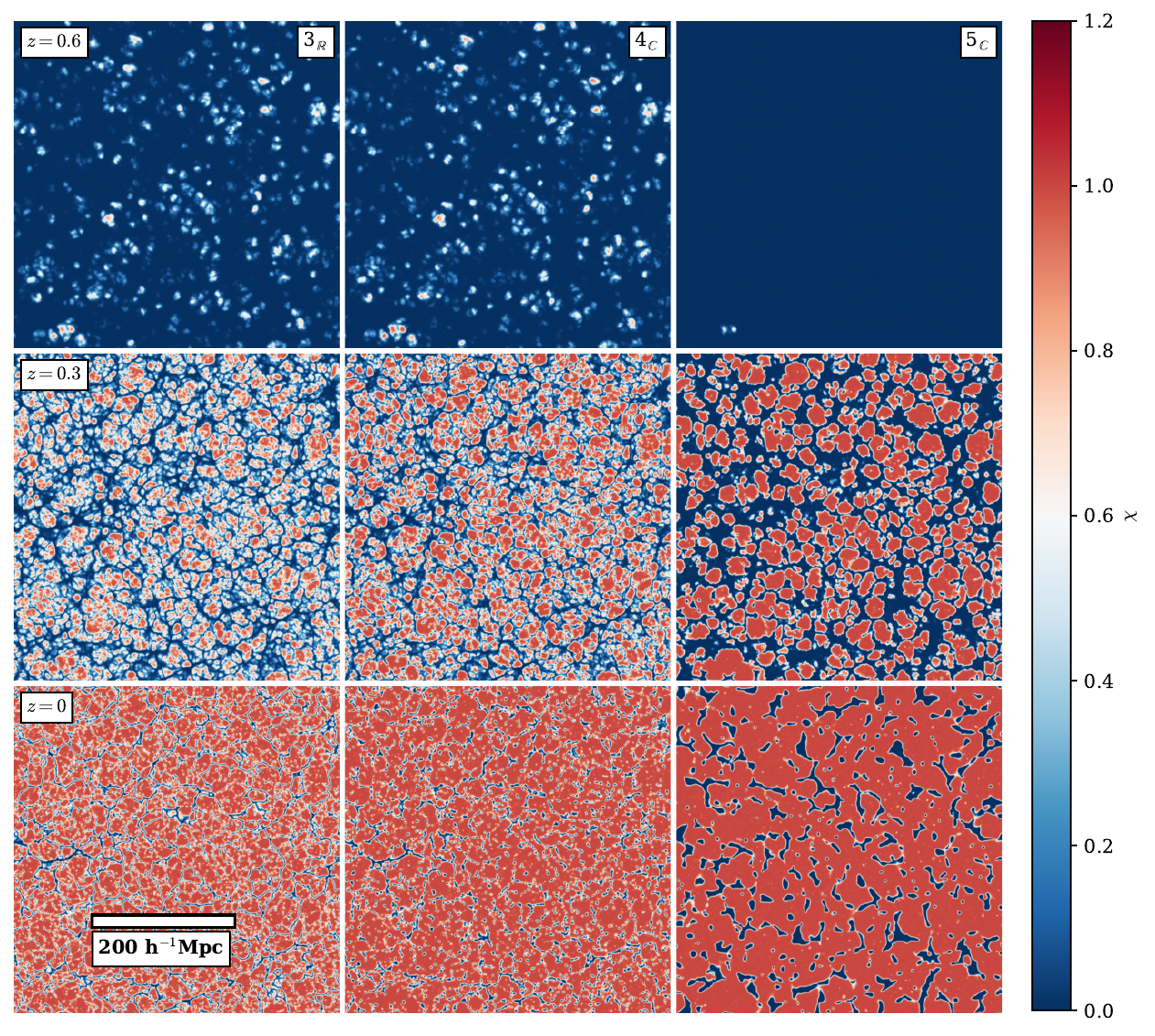
        }
        \caption{The norm of the scalar field $|\chi|$ for different simulations
        and redshifts, shown on a slice of the simulation volume.}
        \label{fig:fieldcartoonA-C}
    \end{figure}

    \paragraph{Field evolution:}
    The norm of the symmetron field is shown on different timeslices of the simulation
    volume for models 1 and 2, both $\phi\in\mathbb{R},\mathbb{C}$, in Fig.~\ref{fig:fieldcartoon1-2},
    while models 3$_{\mathbb{R}}$-5$_{\mathbb{C}}$ are shown in Fig.~\ref{fig:fieldcartoonA-C}.
    Comparing $1_\mathbb{C}$, $2_{\mathbb{C}}$ with $1_{\mathbb{R}}$, $2_{\mathbb{R}}$, we notice an almost
    exact agreement, despite the different field types. At redshift $z=0.6$, the
    phase transition has progressed more in the $\phi\in\mathbb{R}$ case for model
    1. In Fig.~\ref{fig:overview}, we see that this is always the case $\phi\in\mathbb{R}$
    vs.\ $\mathbb{C}$, although it is not so clear in Fig.~\ref{fig:fieldcartoon1-2},
    as cases $\phi\in\mathbb{C}$ also have a steeper rise in the fraction of the
    field in the broken phase, and are mostly caught up by redshift $z=0.6$. At intermediate
    redshift $z=0.3$, $\phi\in\mathbb{C}$ vs.\ $\mathbb{R}$ is almost indistinguishable,
    at which time mostly isolated domains are growing and barely starting to
    intersect, encapsulated by overdense regions. However, for both models 1 and
    2, it can be seen that the $\phi\in\mathbb{R}$ case has a much more clearly
    defined cosmic-web-like structure imprinted on it, mostly in places where it
    is also visible for the $\phi\in\mathbb{C}$ models, indicating the presence
    of overdensities, and a more effectively screened real-valued scalar field.
    In some cases the real-valued field has pinned regions where the complex
    field has none, indicating the presence of domain walls. At redshift $z=0$, we
    now see clearly the overdense regions imprinted, with some filaments showing
    thicker blue arms, likely also representing defects. For the string case, it
    is more subtle as the defects are 1-dimensional and may extend into the plane.

    In Fig.~\ref{fig:fieldcartoonA-C}, showing models 3$_{\mathbb{R}}$-5$_{\mathbb{C}}$,
    we no longer compare analogue $\phi\in\mathbb{R},\mathbb{C}$ simulations. Instead,
    we are showing a more strongly screened case with a smaller self-interaction
    range $L_{\mathrm{C}}=0.45\,h^{-1}\,\mathrm{Mpc}$. Although the phase transition
    still occurs\footnote{Figure~\ref{fig:overview} shows several models that
    have $10^{-5}$-$10^{-3}$ of the volume in the broken phase since $z\sim 0.8$-$0
    .9$.} around $z=0.6$, it is now more progressed in those initial patches,
    with the encapsulating overdensities to the voids where the initial phase
    transitions occur, successfully balancing the gradient energy pushing the field
    to expand, even as the field is fully transitioned $\chi\sim 1$ at $z=0.6$. This
    is also the case for $z=0.3$, where the island domains show $\chi\sim 1$, while
    in model $1_{\mathbb{C}}$ it was $\chi\sim 0.6$. Finally, at redshift $z=0$,
    the domains still successfully expand to cover the entire volume except for
    overdense filaments and defects. In model $5_{\mathbb{C}}$ that has the strongest
    screening with $z_{*}=-1/3$, there are still significant pockets of unbroken
    volume at $z=0$. For smaller $L_{\mathrm{C}}$ and $z_{*}$ still, we expect
    to prolong the phase transition process further, allowing isolated islands of unscreened domains
    to persist longer. This can be seen in Fig.~\ref{fig:overview}, when viewing
    $\phi\in\mathbb{R}, \mathbb{C}$ separately.

    \begin{figure}[ht!]
        \centering
        \includegraphics[width=0.98\linewidth]{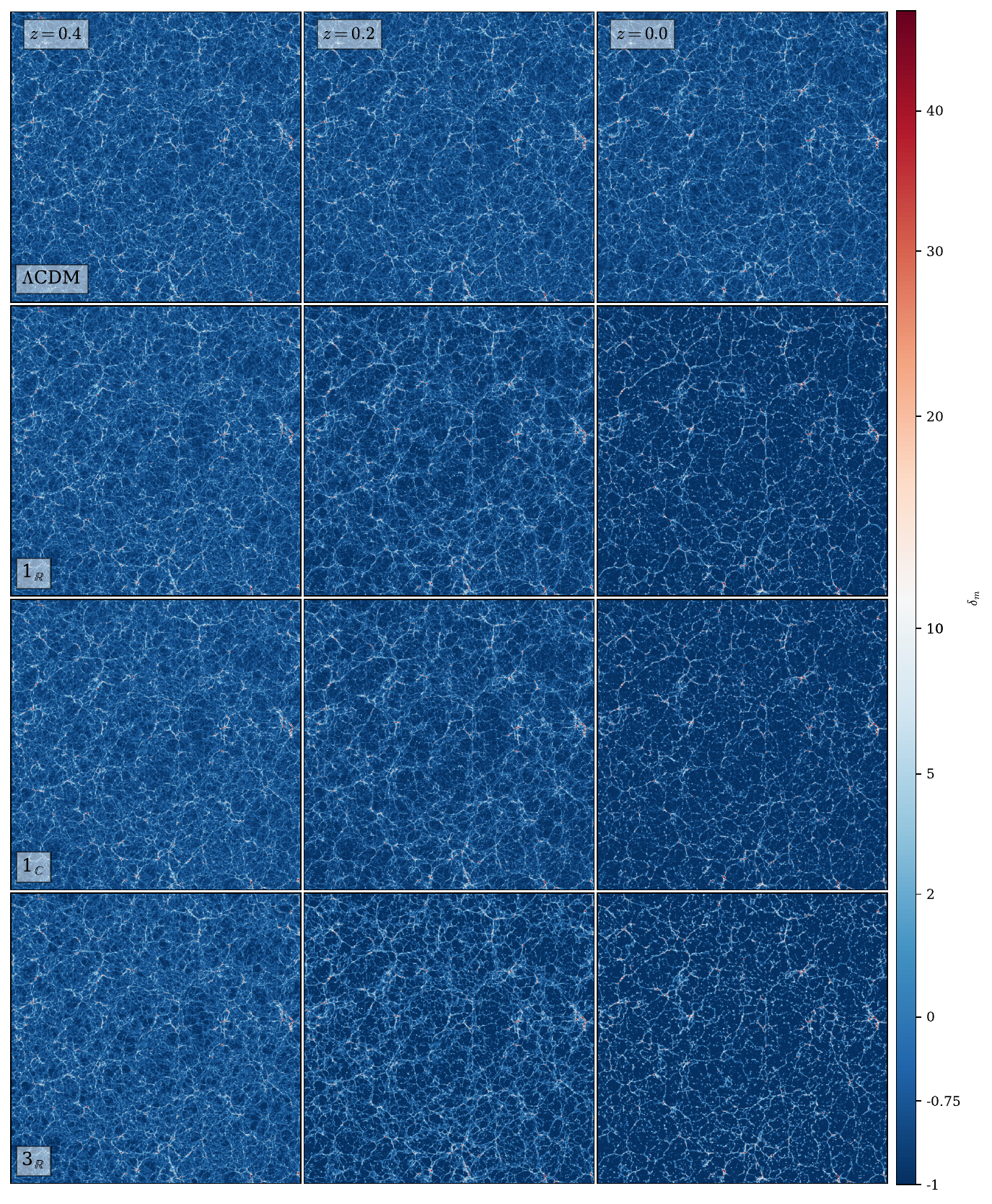}
        \caption{The matter overdensity $\delta_{m}$ for 3 redshifts
        $z=0.4,0.2,0$ for a selection of the simulation suite.}
        \label{fig:densityAll}
    \end{figure}

    \subsection{Density distribution}
    \label{SS:densityDistribution}

    \paragraph{Overview:}
    Figure~\ref{fig:densityAll} shows slices of the matter overdensity field at the
    different redshifts $z=0.4, 0.2, 0$. For brevity, we show the least extreme cases
    here, but animations for all simulations are available at the link in
    footnote \ref{footnote:animations}. At $z=0.4$, it is difficult to tell a
    difference between models $1$ and $\Lambda$CDM, while in model
    $3_{\mathbb{R}}$, subtle enhanced density contrasts can be seen in the underdensities,
    coinciding with the initial areas of phase transition of Fig.~\ref{fig:fieldcartoonA-C}.
    At redshift $z=0.2$ the difference from $\Lambda$CDM is stark, hollowing out
    the underdense regions in between filaments significantly by redshift
    $z=0$, with overdensities being more concentrated in the cosmic web. The
    cosmic web itself seems to be less affected and can easily be compared to
    $\Lambda$CDM, though there are some smaller contrasts there. We can see a few
    filaments that are not visibly present in the $\Lambda$CDM case, especially in
    the $\phi\in\mathbb{R}$ case, likely created by defects. A rich defect network,
    branching on smaller scales for smaller $L_{\mathrm{C}}$ (as can be seen in Fig.~\ref{fig:defect_renderings})
    is thus a means of circumventing the hollowing out of underdensities by these
    strong force strong screening phase transitions. The visibly clear effects
    in Fig.~\ref{fig:densityAll} strongly suggest that the parameter choices made
    for the simulation suite are constrainable by available datasets. Although
    our findings below indicate that they are, the signal strength might be weaker
    than expected from the figure.

    \paragraph{Matter structure:}
    We show the relative differences with respect to $\Lambda$CDM of the matter power
    spectra in Fig.~\ref{fig:pk_matter}. The parameter choices were made to
    have an effect on the matter power spectra $\lesssim 10\%$, which is approximately
    the case, except for simulation $5_{\mathbb{C}}$ where it was difficult to
    predict the impact, see the discussion in Section \ref{SS:parameters}. Again,
    simulations $1, 2$, $\phi\in\mathbb{C}$ vs.\ $\mathbb{R}$ are difficult to separate,
    indicating that the main effect on the matter clustering statistics is not
    coming from defects, but from the screening profiles of the scalar field as set up by
    the environment. However, the enhancements found in models $1, 2$ for $\phi\in
    \mathbb{C}$ are $7.3\%$ and $19\%$ larger, respectively, than in their real-valued
    analogues. This may be due to defects, which in the $\phi\in\mathbb{R}$ case
    more effectively screens larger volumes, as can be seen by looking at the
    defect fraction of the box volume in Figs.~\ref{fig:overview} and \ref{fig:defect_renderings} or the thicker
    regions of $\phi\sim 0$  around the filaments in Fig.~\ref{fig:fieldcartoon1-2}.
    The enhancement spectra are all initially very peaked at a scale $k_*$. This scale
    is seen to vary with $L_{\mathrm{C}}$ between simulations $1, 2$ and $3_{\mathbb{R}}$,
    and scale with the scale factor as a function of the redshift for the
    individual simulations, since $L_{\mathrm{C}}$ is a physical scale, while
    $k$ is the comoving wave number. Presumably, this is related to the initial
    domain size and the region over which $\phi$ is mostly coherent. We find the
    scale to be well approximated by $k_{*}\simeq (5 L_{\mathrm{C}})^{-1}(a_{*}/0
    .909)^{-2}$; it is not clear whether the scaling will persist for a wider
    range of parameters. The high-$k$ growth of power in Fig.~\ref{fig:pk_matter}
    is a resolution effect as we approach the Nyquist frequency at
    $8\,h\,\mathrm{Mpc}^{-1}$. The drop in power before this relates to efficient
    screening at small scales and in the case of simulation $5_{\mathbb{C}}$ in particular
    because of the destruction of small-scale structures from the strong forces and
    matter flows.

    \begin{figure}[tb]
        \centering
        \includegraphics[width=\linewidth]{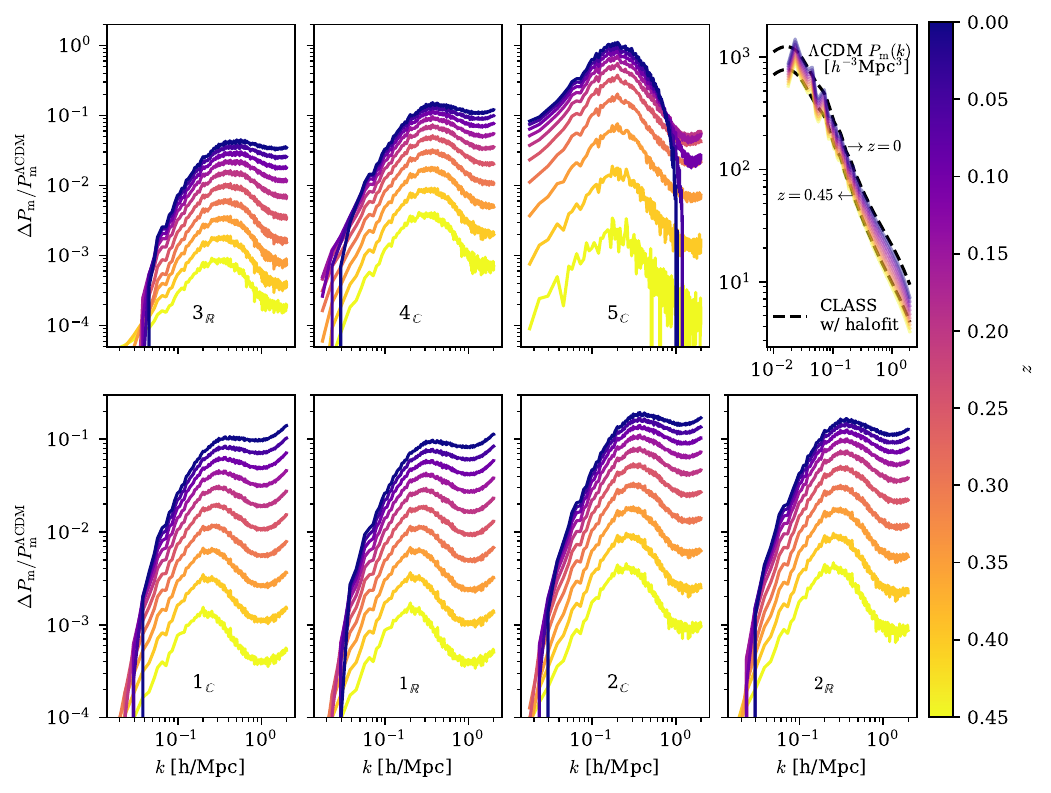}
        \caption{Relative differences in matter power spectra, shown for the different
        simulations in panels 1-4. The colour indicates the redshift, from
        $0.45$ to $0$. The Nyquist frequency is $k_{\mathrm{Nyq}}\approx 8\,h\,\mathrm{Mpc}
        ^{-1}$. All of the spectra show enhancement with respect to the
        $\Lambda$CDM case. Top right panel: The $\Lambda$CDM power spectra $P_{\mathrm{m}}(k)$ for the same redshift range, compared to CLASS with halofit at $z=0.45$ and $z=0$.} 
        \label{fig:pk_matter}
    \end{figure}

    \paragraph{Matter velocities:}
    In Fig.~\ref{fig:CDMvelocities} we show modified velocity statistics of
    matter particles as a consequence of being accelerated by the fifth force. Error
    bars are generated from the variances found in delete-one jackknife sampling,
    on sub-volumes created by splitting the simulation box into 16 equal
    rectangular rods. The bars show 95\% error intervals. We see a clear growth
    of a large velocity tail, larger mean, and log-normal appearance for all simulations.
    Simulation $5_{\mathbb{C}}$ in particular forms a second peak at larger
    velocities that has a larger area than the original peak by redshift $z=0$.

    \begin{figure}[tb]
        \centering
        \includegraphics[width=\linewidth]{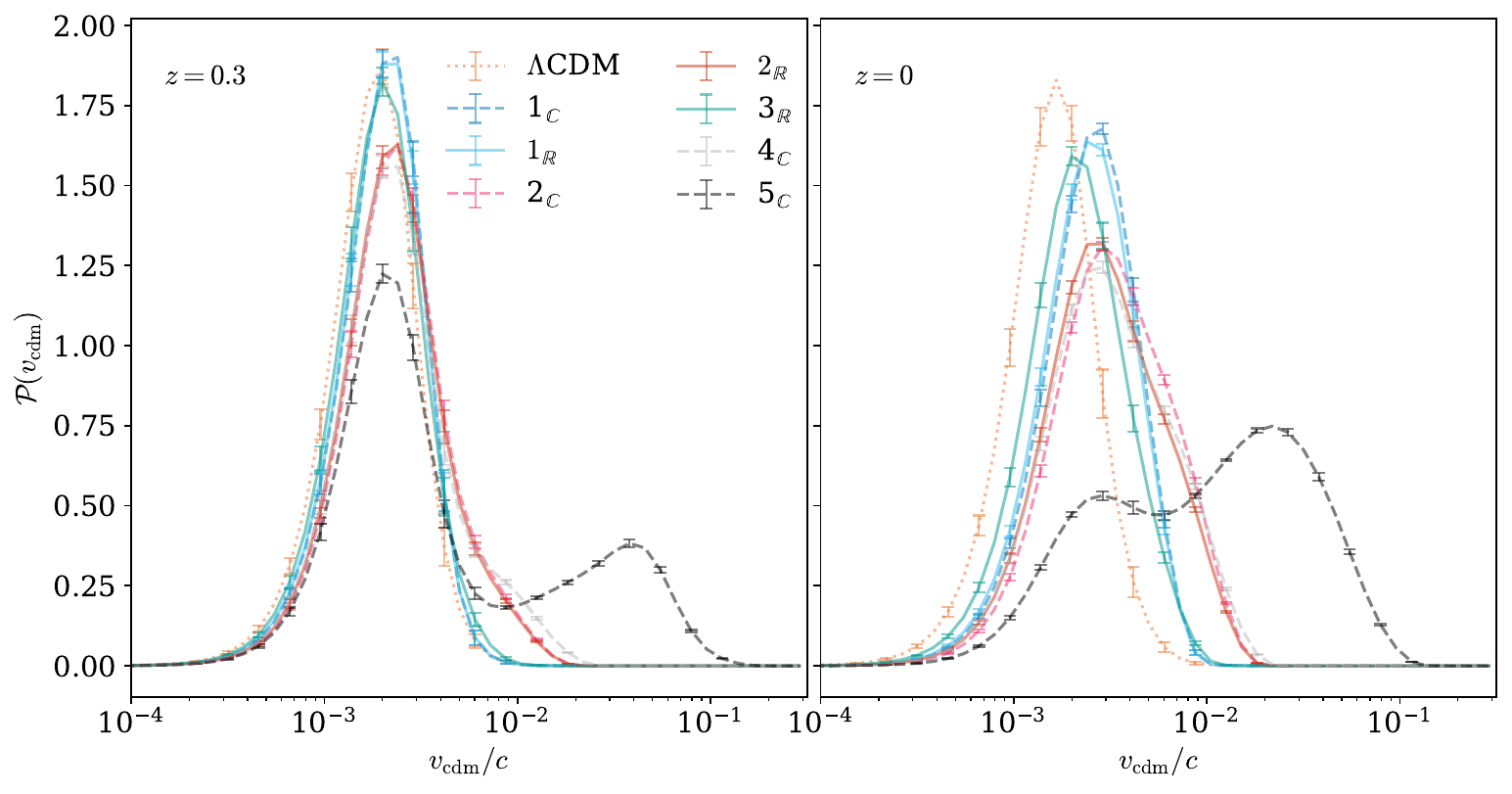}
        \caption{Velocity histograms of the cold dark matter particles at
        different redshifts, normalised to probability distributions. Error bars
        show $2\,\sigma$ intervals in the jackknife scatter. Logarithmic scale is
        chosen to highlight the appearance of a second peak in model $5_{\mathbb{C}}$.
        }
        \label{fig:CDMvelocities}
    \end{figure}

    \paragraph{Nonlinear structures:}
    Figure~\ref{fig:densityAll} described in the first paragraph of Section
    \ref{SS:densityDistribution} shows large relative effects on the matter
    distribution in the underdense environments for the non-$\Lambda$CDM choices,
    while the overdense filaments are mostly intact. Here we investigate the effect
    on virialised structures or matter halos, which predominantly form in
    overdense environments. The \texttt{Rockstar} halo finder \citep{behroozi_rockstar_2013}
    identifies halos in simulation snapshots by constructing phase-space trees of
    gravitationally bound particles. The gravitational binding used in \texttt{rockstar} is found assuming $\Lambda$CDM using the standard rockstar halo definition, which comes with some caveats and should be modified in a more complete approach. After compiling the full halo catalogues, we
    remove subhalos bound to parent halos from our analysis. We use the mass
    definition $m_{200\rm b}$ that defines the mass of the halo as the mass
    enclosed within the largest spherical shell within which the average matter
    density is 200 times greater than the background density. The halo finder finds
    halos with more than 50 particles while putting the force resolution to $\mathrm{d x}_{\mathrm{F}}=0.06\;h^{-1}\mathrm{Mpc}$, which is smaller than the one suggested by the grid spacing by a factor of $4$
    in order to have sufficiently large catalogues, with the caveat that some of
    the smaller halos can be a result of noise. For now, we are looking at
    summary statistics and not substructure, which allows less conservative resolution
    settings. We select the $N=500\,000$ most massive halos from each halo catalogue
    for the analysis, except for the halo mass functions, where we keep all halos.
    In comparing different simulations, fixing the number of halos is the more observationally
    relevant comparison, and is the closer analogue to a survey where they would observe a
    number of galaxies limited by their luminosities. 

    \begin{figure}[tb]
        \centering
        \includegraphics[width=\linewidth]{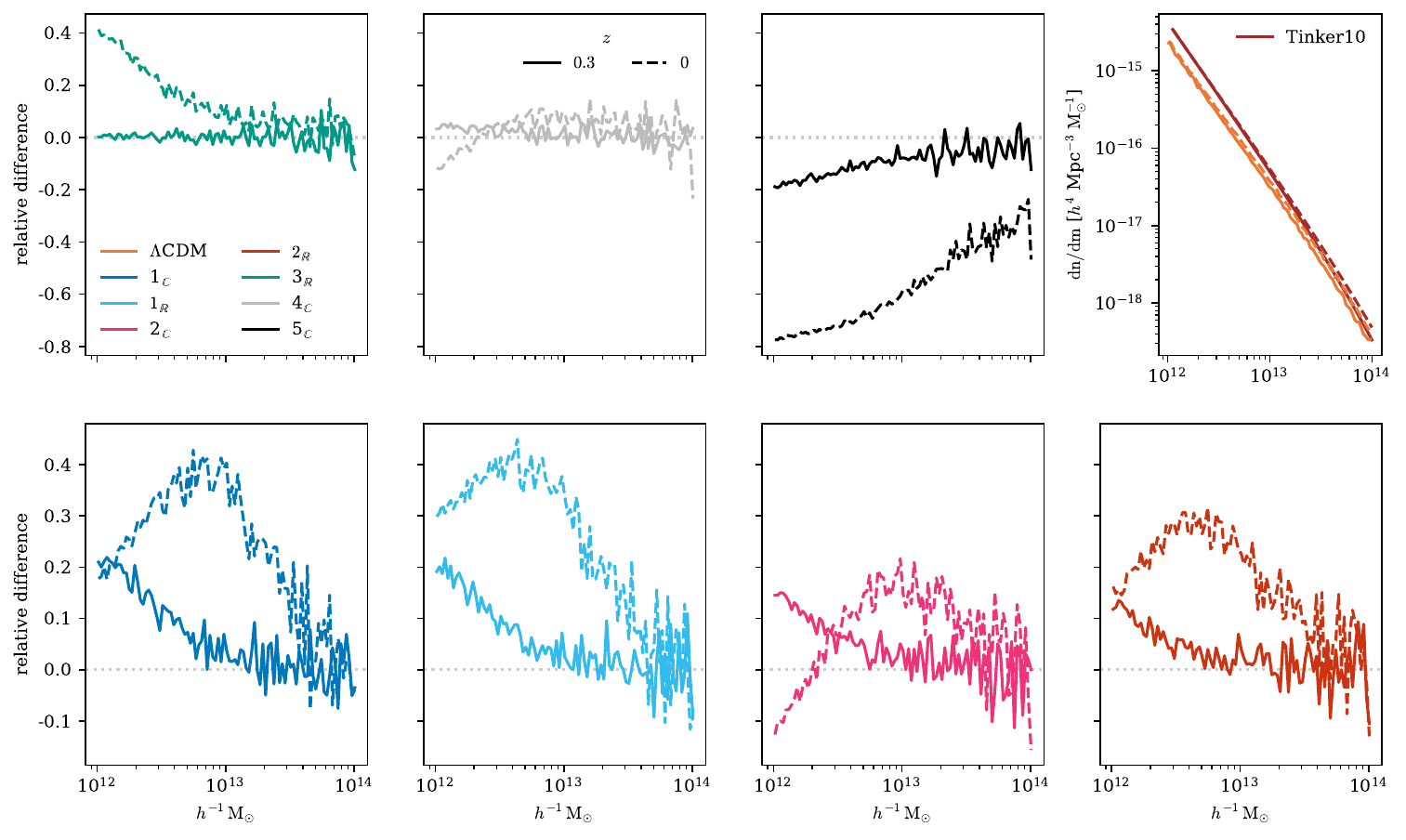}
        \caption{Top: Halo mass functions $\mathrm{d}n/\mathrm{d}m$ for different
        simulations (colour) and redshifts (linestyle), as a function of the
        solar mass $\mathrm{M}_{\odot}$. Bottom: Relative differences with
        respect to the $\Lambda$CDM case. Halos with masses
        $M\ll 10^{13}\,h^{-1}\,\mathrm{M}_{\odot}$ are poorly resolved
        and should be interpreted carefully.}
        \label{fig:HMFoverview}
    \end{figure}

    \paragraph{Halo mass function:}
    Figure~\ref{fig:HMFoverview} shows the Halo Mass Functions (HMFs) of the halo
    catalogues at redshifts $z=0.3$ and $z=0$. The mass function is shown and
    compared to the parametrisation of Tinker et al. \citep{Tinker:2008ff}
    in the case of $\Lambda$CDM, while each of the other simulations is given in
    terms of relative difference with respect to the simulation of $\Lambda$CDM.
    There is a clearly defined peak in the HMF enhancement, where the abundance
    of halos is significantly enhanced with respect to $\Lambda$CDM, in most scenarios.
    This peak appears to shift towards larger masses with time. Counter-intuitively,
    halo masses smaller than the peak appear to be suppressed rather than
    enhanced, but we only observe this effect in the simulations where $\phi\in\mathbb{C}$.
 Indeed, in all cases, there are smaller enhancements for $\phi\in\mathbb{C}$ than for $\phi\in\mathbb{R}$. This is clear when comparing
    simulation variants of simulations $1, 2$ that differ only in this respect.
    Simulations 3$_{\mathbb{R}}$ and $4_{\mathbb{C}}$ also show this trend, even
    as simulation $4_{\mathbb{C}}$ has twice as large a fifth force strength parameter
    $\beta$. The effect on the HMF of simulation $4_{\mathbb{C}}$
    remains within $\sim 10\%$ of $\Lambda$CDM, despite the more dramatic effects on
    the matter density distribution than seen in Fig.~\ref{fig:densityAll}. Simulation
    $5_{\mathbb{C}}$ is again the most extreme case and shows a suppression
    across all masses.

    Understanding the reason for the suppression in the HMF for model $5_{\mathbb{C}}$ and for low masses for models $2_{\mathbb{C}},4_{\mathbb{C}}$ requires running more careful analysis of the many contributing effects to their formation and migration away from underdense areas. One should also explore the effect of modifying the \texttt{Rockstar} halo binding energy, which in the presence of the fifth force may allow larger velocity particles to bind to a halo. These are interesting future directions that are outside of the scope of the current work.
    The above observations should furthermore
    be taken with the caveat that the small mass halos (less than $\sim 10^{13}\,
    h^{-1}\,\mathrm{M}_\odot$) are poorly resolved, see e.g. the growing mismatch with the Tinker mass function at smaller masses, Fig.~\ref{fig:HMFoverview}. Optimistically, the
    relative differences are less sensitive to resolution effects, but ideally these
    trends should be investigated in simulations with larger dynamical ranges of
    scales and a more conservative halo definition.

    \begin{figure}[tb]
        \centering
        \includegraphics[width=\linewidth]{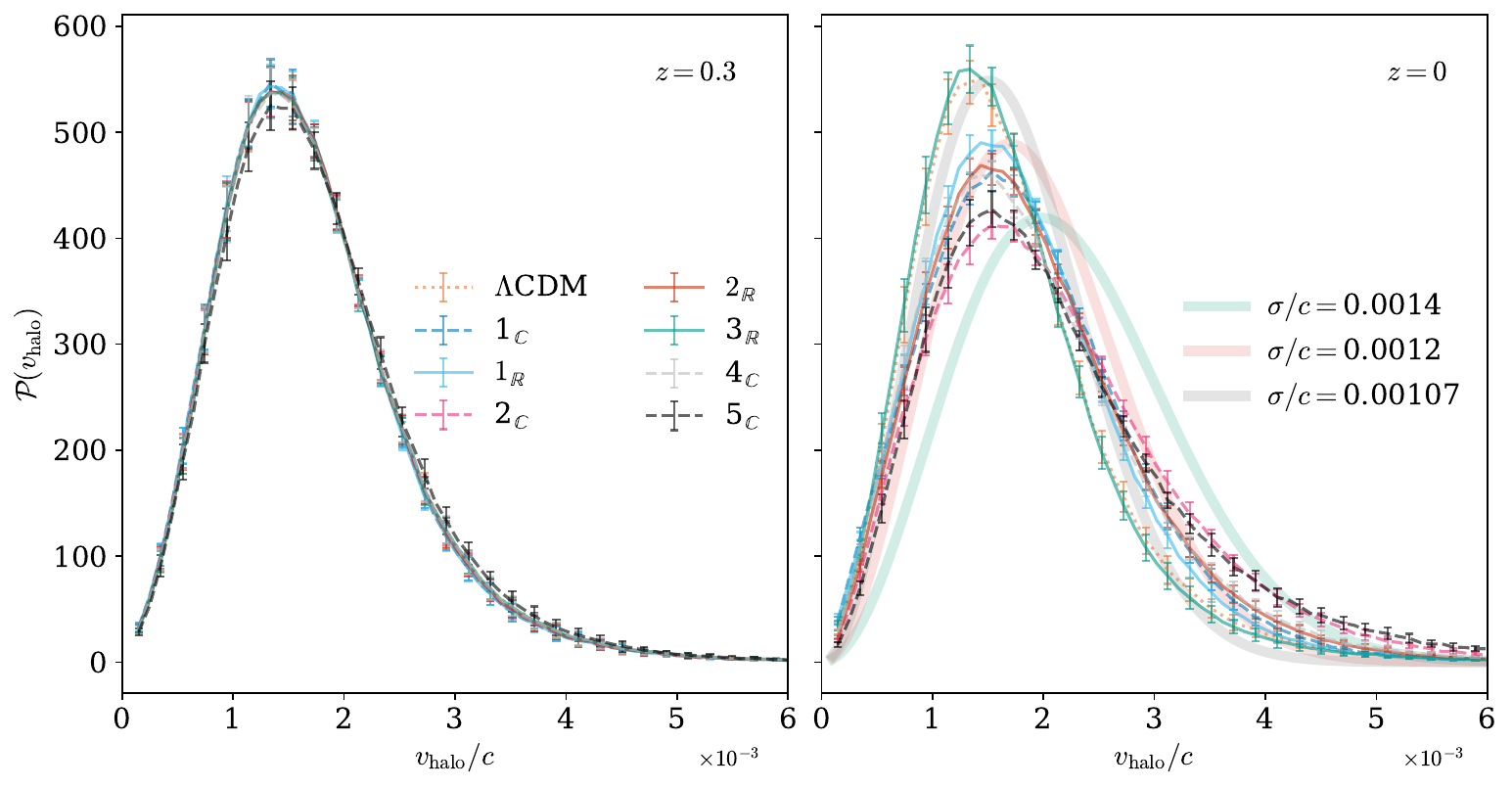}
        \caption{Velocity histograms of $500\,000$ most massive halos halos at
        redshifts $z=0.3, 0$, normalised to probability distributions. Shown
        here for the different simulations. Error bars show $2\,\sigma$ intervals
        in the jackknife scatter. }
        \label{fig:HalosVelocities}
    \end{figure}

    \paragraph{Halo velocities:}
    Figure~\ref{fig:HalosVelocities} shows the velocity distributions of the halo
    catalogues at the different redshifts $z=0.3, 0$, with error bars produced in
    the same way as for the matter velocity distribution, Fig.~\ref{fig:CDMvelocities}.
    At $z=0.3$, the effect on their velocity dispersion is very marginal and
    consistent with each other within 1-2\,$\sigma$ error bars, apart from simulation
    $5_{\mathbb{C}}$ that is barely $2\,\sigma$ lower in amplitude at the peak of
    the distribution. At redshift $z=0$, only simulation 3$_{\mathbb{R}}$ has a
    compatible distribution with $\Lambda$CDM, while the rest follow more tailed
    distributions, with velocities reaching percent of the speed of light. For
    comparison, Maxwell-Boltzmann distributions with different velocity dispersion
    $\sigma_{v}/c$ are overlaid in wide bands, showing that the different
    simulations range from $\sim 12$-$30\%$ larger effective velocity dispersion
    in the halo catalogues, compared to $\Lambda$CDM. This will likely cause observable
    effects in galaxy surveys, where galaxy velocities cause redshift space
    distortions in the reconstructed halo positions, and the appearance of a characteristic
    quadrupole in the effective distribution \citep{Kaiser:1987qv}.

    \subsection{Probes of environment-dependent effects}
    \label{SS:probes}

    In the previous section, we presented $\sim 10$-$20\%$ effects on summary
    statistics such as matter power spectra and velocity histograms (excluding simulation
    $5_{\mathbb{C}}$ for now). The abundance of halos is more affected, with
    effects of up to $\sim 40\%$. However, the halo abundances in simulations 2
    and $4_{\mathbb{C}}$ are affected only at $20\%$ and $10\%$, respectively.
    Figure~\ref{fig:densityAll} shows more dramatic but very environment-dependent
    effects on the density distribution. In this section, we will investigate whether
    environment-dependent probes can capture this signal successfully and thus become
    target probes for constraining SIPT-type models qualitatively similar to the ones presented here.

    \paragraph{Matter density PDF:}
    A Gaussian probability distribution is uniquely defined in terms of its mean
    and variance. The typical scenario in study is that of adiabatically sourced,
    Gaussian initial density perturbations produced by cosmological inflation, and
    subsequently acted upon by linear physics. Power spectra and 2-point
    correlation functions are therefore widely used in cosmological analysis. In
    the late-time Universe nonlinearities grow and cause the density distribution
    to acquire non-Gaussian features \citep{Bernardeau:1992zw}. Recent and upcoming surveys, e.g. Euclid \cite{Euclid:2024yrr}, are therefore increasingly targeting nonlinear statistics such as bispectra or voids distributions. The nonlinearities may also be  accessed by considering the Probability Density Function (PDF) – or equivalently the $1$-point distribution function. Analytic predictions for the PDF can be made using Large Deviations
    Theory (LDT) \citep{Bernardeau:2015khs,Uhlemann:2015npz}. This gives us
    a window into environment-dependent physics by encoding nonlinear physics such as
    the density-dependent screening of the symmetron. We follow the procedure laid
    out in \cite{Cataneo:2021xlx}: first, we smooth the overdensity field from
    our simulations with a top-hat smoothing window; then, we generate the PDF of
    the resulting overdensity field. For comparison, we apply the Python package
    \texttt{pyLDT-cosmo} by \cite{Cataneo:2021xlx}, to which we have to input
    log-density variances found in our simulation snapshots.

    \begin{figure}[tb]
        \centering
        \includegraphics[width=\linewidth]{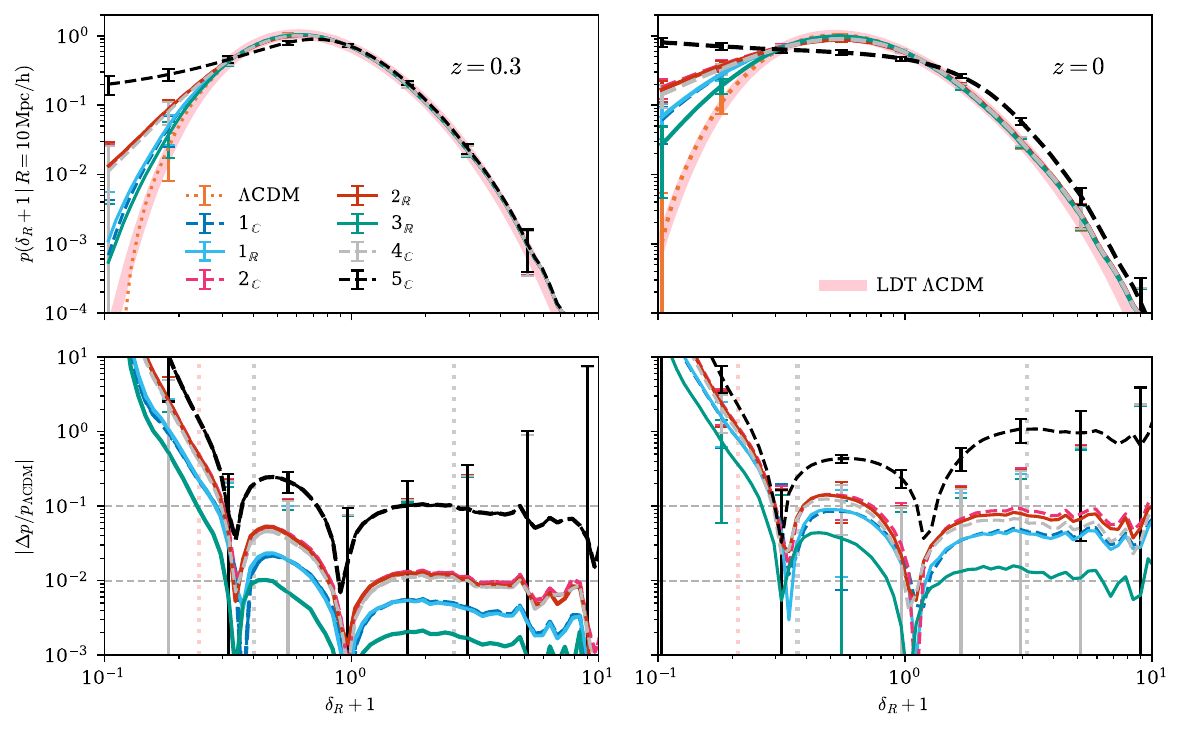}
        \caption{ Density histograms of the density fields of matter for the different
        models and redshifts. The density fields are smoothed with a spherical
        top-hat filter of $R=10\,h^{-1}\,\mathrm{Mpc}$ radius. Top row: The
        density probabilities. Bottom row: The relative differences of the
        density probabilities with respect to the $\Lambda$CDM model case. The
        dotted vertical lines indicate the $1$\textperthousand\,
 (red), $3\%$ and $90\%$ quantiles of the
        respective $\Lambda$CDM CDFs $c(\delta_{R}>x;z)$. 
        Errorbars indicate $2\sigma$ intervals in the Poisson noise.
        }
        \label{fig:hist1pdensity}
    \end{figure}

    In Fig.~\ref{fig:hist1pdensity}, we show the 1-point distribution function, or
    the PDF, for a density field that is smoothed with a top-hat filter with
    smoothing radius $R=10\,h^{-1}\,\mathrm{Mpc}$. The predictions for $\Lambda$CDM
    are made using the variance of the log-density distribution
    $S_{2} \equiv \mathrm{Var}(\log[1+\delta_{m}])$. At redshifts $z=0$ and $z=0.
    3$, the $\Lambda$CDM simulation gives $S_{2}^{z=0}=0.379226$ and $S_{2}^{z=0.3}
    =0.291989$, respectively. Compared with the variance reported for $z=0$ in table
    B1 of \cite{Cataneo:2021xlx}, ours is $3.3\%$ smaller, which we assume is
    due to our different simulation setup and slightly different cosmological parameters.
    We find good agreement with the LDT prediction for the $\Lambda$CDM simulation. The small differences may be due to our particle-mesh-based code, compared
    to the Quijote simulation suite used in \cite{Cataneo:2021xlx} which
    uses the tree-PM force solver of \texttt{Gadget-3}, in turn based on \texttt{Gadget-2}
    \citep{Springel:2005mi}. Additionally, low probabilities
    $p\ll 10\%$ are increasingly subject to sample variances from the finite
    simulation volume. We indicate the sample variance of the limited number, $N$, of independent spheres of the smoothing radius inside of the box volume per bin $\Delta\delta_R$, with the formula $\sigma_{\mathrm{PDF}}= \sqrt{n_i}/(N_{\mathrm{sphere}}\;\Delta_{ \delta_{R}})$, where $n_i$ is the bin count of the $i$th simulation. We approximate $N_{\mathrm{sphere}}$ by the volume fraction $N_{\mathrm{sphere}}\approx V_{\mathrm{box}}/V_{\mathrm{sphere}}$. Relative differences to $\Lambda$CDM ($i=0$) are estimated using the formula $\sigma_{\mathrm{rel}} = \sqrt{n_i/n_0^2+n_i^2/n_0^3}$, which can be derived from Eq.~(8) in \cite{diaz-frances_existence_2013} after inserting $\mu_i=n_i$ and $\sigma_i=\sqrt{n_i}$ for the mean and standard deviation. Compared to our $\Lambda$CDM simulation, the PDFs of the different models
    1-2 and 3-5 all have a large enhancement in underdensities $\delta_{R}\lesssim
    -0.65$, reaching $100\%$ for $z=0$ at $\delta_{R}\sim -0.75,-0.9$ depending
    on the model, and reaching a factor $\sim 10$ by $\delta_{R}\lesssim -0.85$. A pattern of enhancement can be seen at small and large densities, and suppression at intermediate densities $-0.7\lesssim \delta_R \lesssim 0$. We note that the signal is too small to beat sample variance for the relative differences at $z=0.3$ (apart from for simulation $5_{\mathbb{C}}$), but it becomes significant for $\delta_R\lesssim -0.3$ at $z=0$. 

    \paragraph{PDF detectability:}
    In Section 5.6 of \cite{Uhlemann:2019gni} a short discussion of applicability
    to survey data is provided. An option is to find the weak-lensing
    convergence PDF in tomographic redshift slices, as was done in \cite{Barthelemy:2019ciu}.
    As discussed there, this is in particular relevant to the Euclid survey \citep{Euclid:2024yrr}
    that will provide accurate weak-lensing measurements over half the age of
    the Universe. This is less relevant for our case as the Euclid survey will cover
    redshifts $z=$0.9-1.8. Instead, e.g. SDSS data (total redshift range $z=0$-$0.7$) can be applied; this was done with count-in-cell PDFs of galaxies in \cite{Yang:2010qs}. Finally, `density-split statistics' combines the
    two approaches to relate the galaxy count PDF to the matter density PDF. The
    Fisher analysis in \cite{Uhlemann:2019gni} was made directly on the
    matter PDFs, discarding the 3\% smallest densities and 10\% largest densities,
    as they found them to be the most sensitive to resolution effects. The $3$ and
    $90$ percentiles of the Cumulative Distribution Functions (CDFs) are indicated
    in the lower panels of Fig.~\ref{fig:hist1pdensity}, showing that the main signal
    of our models would thus be discarded in this analysis. For the percentiles
    that would be included, the signal of our models is $\sim 0.1$-$1\%$ and $1$-$1
    0\%$
    at redshifts $z=0.3, 0$, respectively, which is greater than the effects of
    $\sim 2\%$ for the massive neutrino models shown in Fig.~10 of \cite{Uhlemann:2019gni}.
    By combining three redshifts, three smoothing radii, and the matter power
    spectra up to quasilinear scales, they claim a BOSS-like survey ($z=$0.15-0.7)
    can put a $5\sigma$ lower constraint on the sum of neutrino masses, which
    is impacting the matter PDF at $\sim 1$-$2\%$. Although these late-time
    phase transition (SIPT) models are active for a shorter time duration and thus
    benefit less from the boosted signal and the removal of degeneracies by the
    use of multiple redshifts, it is still likely that they are distinguishable from
    $\Lambda$CDM by using their large signal at very small redshifts.
    Extending the
    analysis to include lower percentile density data will significantly boost the
    signal, but this is more subject to sample variance as the tails of the PDF correspond to rarer events. 
    The DESI Bright Galaxy Survey (BGS) catalogue covers the redshift range $0.1 < z < 0.3$.
    A lower bound on the sample variance can be made by counting the number, $N_{\mathrm{sphere}}$, of $R=10\;h^{-1}\;\mathrm{Mpc}$ spheres in the corresponding comoving survey volume (which is greater than our simulation box). 
    Ignoring masking and sky coverage, we find $N_{\mathrm{sphere}} \sim 5.6\times 10^5$, corresponding to a Poisson relative error of $\Delta p/p\sim \pm 10\%$ at 2$\sigma$ for a $1$\textperthousand\, quantile (smallest quartile is shown by red vertical line in Fig.~\ref{fig:hist1pdensity}). This comes on top of the uncertainties related to tracer sampling and systematics that can be significant. However, the signal is $\mathcal{O}(1)$, which is likely to be sufficiently large for detection.
    While interesting, a full Fisher analysis forecast of detectability of
    these structure induced symmetron models is beyond the scope of this paper.

    \begin{figure}[tb]
        \centering
        \includegraphics[width=0.9\linewidth]{
            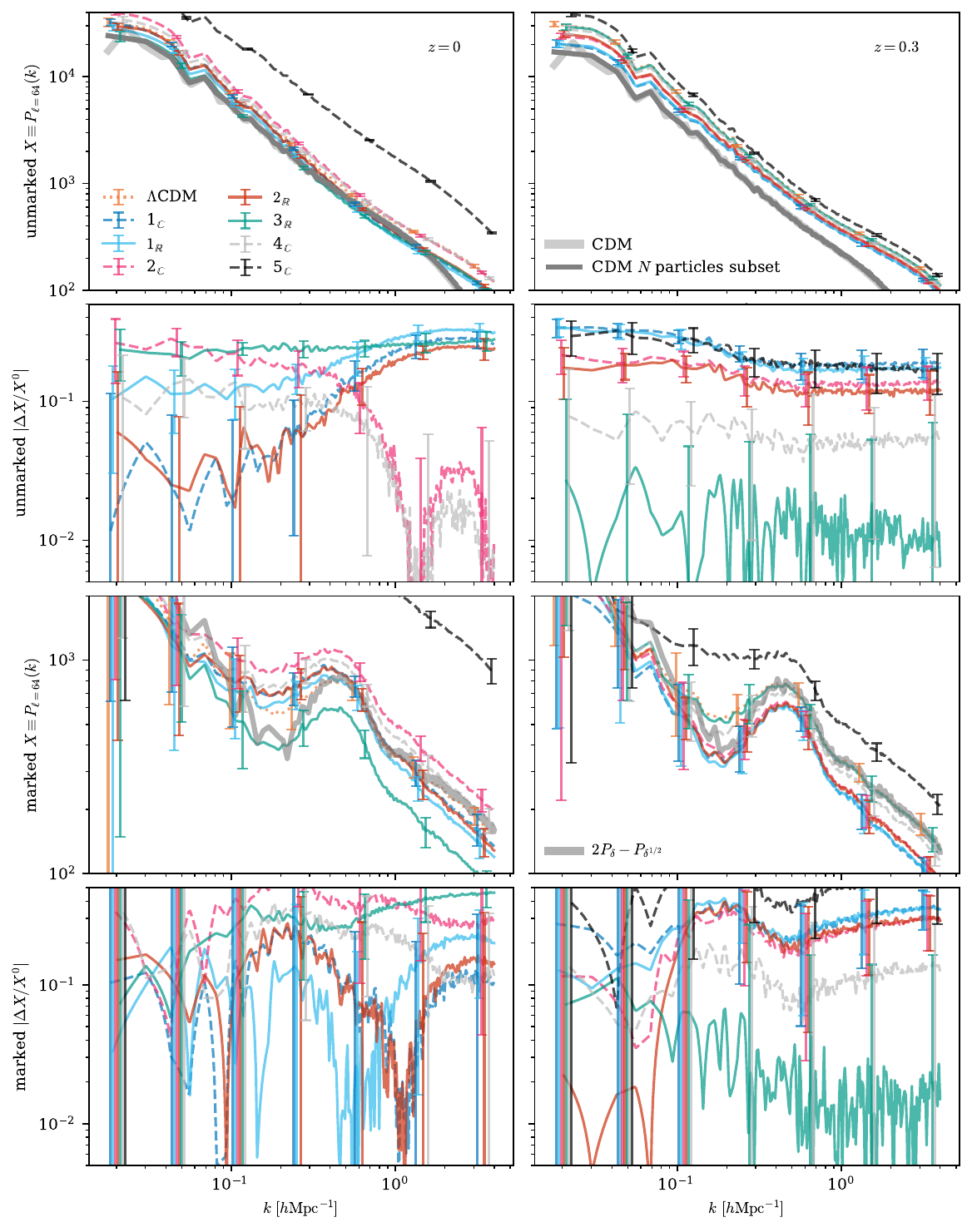
        }
        \caption{Halo power spectra for the different models at redshifts $z=0$
        and $z=0.3$. Errorbars represent $2\,\sigma$ jackknife scatter. Top row: Monopole power spectra of the halo catalogues in
        the different models.
        The thick
        solid grey lines show the spectra computed from \emph{all} particles and
        from a random subset of $500\,000$ particles, respectively. 
        Second row: Absolute relative differences
        of the power spectra between the models at the two redshifts, with respect
        to the $\Lambda$CDM case. Bottom rows: Same as top rows, but for the marked halo power spectra, using the mark
        defined in Eq.~\eqref{eq:markchoice}. 
        }
        \label{fig:halopowerspecsWmarks}
    \end{figure}

    \paragraph{Halo spectra and marked statistics:}
    A second approach to boosting the signal strength for environment-dependent physics
    is using marked power spectra \cite{stoyan_correlations_1984}.
    In this approach we are finding the summary statistics of a field that is transformed (marked) as $\delta_{\mathrm{m}}\rightarrow m(\delta_{\mathrm{m}},\dots) \delta_{\mathrm{m}}$. In \cite{White:2016yhs} a specific mark was suggested for probing
    environment-dependent screening models of modified gravity
    \begin{equation}
        \label{eq:markchoice}m(x; R,p,\delta_{s}) = \left(1+ \frac{\delta_{R}(x)}{1+\delta_{s}}
        \right)^{-p},
    \end{equation}
    where $\delta_{R}$ again is the density field smoothed with a top-hat filter
    over a radius $R$, $\delta_{s}$ parametrises the sensitivity of the mark to
    the local density and larger exponents $|p|$ give more environmentally
    sensitive weights. Positive $p$ are up-weighting underdense environments,
    which is beneficial in our case as this is where the most dramatic effects
    occur; see Fig.~\ref{fig:densityAll}. In \cite{Massara:2020pli} 125 different
    choices of the parameters $R, \delta_{s}, p$ were considered for the case of
    massive neutrinos, and the choice $R=10\,h^{-1}\,\mathrm{Mpc}$, $p=2$ and $\delta
    _{s}=0.25$ was found to give the tightest constraints on cosmological parameters.
    An exhaustive study of optimal weight choices is beyond the scope of our
    work. Instead, as a proof of concept, we will adopt the mark \eqref{eq:markchoice}
    with parameters $R=10\,h^{-1}\,\mathrm{Mpc}$, $p=2$, $\delta_{s}=0.25$.

    Both marked and umarked halo density power spectra are shown in Fig.~\ref{fig:halopowerspecsWmarks}.
    They are found, as described in the start of this subsection, from the
    $N_{\mathrm{h}}=500\,000$ most massive halos in the simulation. For each
    unmarked power spectrum, there is expected a shot noise contribution due to the limited tracer number $P_{\rm{SN}}=V/N_{\mathrm{h}}$, where $V$ is the simulation volume. We are here removing this by splitting the halo population into two random subsamples that we cross-correlate. Their respective shot-noise contributions will be uncorrelated and thus disappear in the cross-correlation. This agrees with the result we get by subtracting the shot noise from the full population auto-correlation.
    As an additional check, we find good agreement between the full set of matter particles'
    power spectrum and the spectrum found by randomly selecting
    $N=N_{\mathrm{h}}$ matter particles and performing the cross-correlation procedure on this sample. In the marked case,
    the marks change the noise properties of the spectra. Assuming
    independence of the weights and tracers one can find the analytic
    generalisation of the shot noise contribution
    $P_{\rm{SN}}=V \langle m^{2}\rangle/(\langle m\rangle^{2} N_{\mathrm{h}})$
    \citep{Philcox:2020fqx}. In reality, the marks and tracers are highly correlated
    by design, which can introduce scale-dependent shot-noise and require a more
    careful modelling. We again do the split-population cross-correlation in the marked case to remove any linear contributions from the shot noise. The remaining noise is estimated from the variance found in 16 delete-1 jackknife samples. As an indication of residual shot noise level, we plot for $\Lambda$CDM the spectrum $2P_\delta-P_{\delta^{1/2}}$, where $P_\delta$ is the auto-spectrum of the full population and $P_{\delta^{1/2}}$ that of half of the population. This should subtract the shot noise contribution in the case where it is predominantly a linear addition, and it agrees with the cross-correlation result within the jackknife-estimated error. There is a particular good agreement on large $k$, but deviations of size of the overall signals on smaller $k\lesssim 5\times 10^{-1}$. Indeed, the jackknife errors indicate that the signal to noise vanishes in this range. As such, these scales may be better explored in larger halo catalogs $N_\mathrm{h}\gg 5\times 10^5$, or with better nonlinear noise treatment. 

    The $\Lambda$CDM halo catalogue has a linear bias $b_{0}\approx 1.07$
    compared to the matter density field, when matched in the unmarked spectra on scales $k<0.4\,h\,\mathrm{Mpc}
    ^{-1}$ at $z=0$. The population at $z=0.3$ has a linear bias
    $b_{0.3}\approx 1.27$.
    In general, we see a greater sensitivity to the SIPT-type
    symmetron in the halo spectra relative to $\Lambda$CDM than in the matter
    power spectra, Fig.~\ref{fig:pk_matter}. The unmarked halo spectra, Fig.~\ref{fig:halopowerspecsWmarks},
    show both relative enhancements and suppressions. Model 3$_{\mathbb{R}}$, which
    has the weakest signal in the matter power spectrum, has one of the strongest
    overall signals in the halo spectrum at $z=0$, showing a nearly scale-invariant
    suppression at 20-25\%, while being sub-percent at redshift $z=0.3$. Other
    models, such as $1, 2$, are expressed most strongly on large scales at
    $z=0.3$, while peaking on small scales at $z=0$. model $4_{\mathbb{C}}$
    stands out as a model that is less strongly expressed in the halos than the
    matter particles, staying below 10\% here, while having matter power spectra
    enhanced at $\sim 15 \%$. All models, apart from $5_{\mathbb{C}}$, show
    initial suppression of halo clustering at redshift $z=0.3$, while $2_{\mathbb{C}}$
    and $4_{\mathbb{C}}$ have enhanced halo clustering by $z=0$. The marked
    spectra are mostly successfully boosting the signal at scales close to
    $k\sim \pi/(2 R)\sim 0.15\,h\,\mathrm{Mpc}^{-1}$, while sometimes weakening the signal at smaller
    and larger scales. The
    improved signal strength at large scales in the marked case at $z=0.3$, is a
    factor $\sim 5$-$10$ for model $3_{\mathbb{R}}$, while more modest for the rest. At
    redshift $z=0$, model 1$_{\mathbb{C}}$ and model $2_{\mathbb{R}}$ are improving by a factor $\sim 10$ at large scales, while e.g. models $3_{\mathbb{R}}$ and $1_{\mathbb{R}}$ seem overall worse. Unfortunately, most of the gain of the marked cases is lost in the larger jackknife errors, with the unmarked spectra performing better overall when error is included. This is likely to improve on the use of larger tracer catalogues.

    \paragraph{Halo spectra detectability:}
    Matter halos are not directly accessible in observations, but are hosting
    clusters of galaxies that are. Connecting the effects on the halo statistics
    to those of accessible tracers like galaxies requires an abundance matching scheme,
    see, e.g. \cite{Stiskalek:2021ckt}. For simplification, here we
    did a mass-ranked abundance matching, where we assume that the $N$ largest mass dark matter halos are hosting the $N$ brightest central galaxies that would be detected in a magnitude-limited galaxy survey, dismissing e.g. satellite galaxies. Galaxy survey catalogues are composed of galaxies at different redshifts
    according to radial distance, which has the potential to dilute the overall
    signal for wide redshift bins and models that go from halo clustering power
    suppression to enhancement, such as $2_{\mathbb{C}}$ and 4$_{\mathbb{C}}$.
    Smaller redshifts, where we are seeing stronger effects, are limited by
    small observational survey volumes. Although the Euclid survey is expected to
    have sub-percent errors on the galaxy spectrum measurements at scales smaller
    than $k\sim 0.45\,h\,\mathrm{Mpc}^{-1}$, it will only take data between
    $z=0.9$-$1.8$ \citep{Euclid:2024yrr}, within which our current
    selection of models mimics $\Lambda$CDM, apart from possible foreground
    effects. The BOSS galaxy survey, see, e.g. \citep{Philcox:2021kcw}, covers
    smaller redshifts $z=0.15$-$0.7$, and the DESI survey has galaxy catalogues
    starting at $z\sim 0.1$. In \citep{DESI:2024hhd}, the DESI full-shape analysis
    was performed in 6 redshift bins, where the magnitude-limited BGS has $0.1<z<0.4$, and the lowest redshift bin of Luminous Red Galaxies
    (LRG) has $0.4<z<0.6$. The rest of the bins have $z>0.6$ and would thus be insensitive to
    this selection of models. Taking into account the reduced data in the
    lower redshift, it is still likely that the DESI and SDSS data can detect
    deviations from $\Lambda$CDM: At large scales ($k\lesssim  0.1 \;h\,\mathrm{Mpc}^{-1}$), the DESI BGS has a similar number of tracers to us ($N_\mathrm t\sim  3 \times 10^ 5 $) and an effective redshift $z_{\mathrm{eff}}\sim 0.3$. Figure 17 of \cite{DESI:2024jxi} shows that the statistical and systematic $2\;\sigma$ errors are roughly $\pm 5 \%$ at $k\sim0.1 \;h\,\mathrm{Mpc}^{-1} $, which can be further reduced by using larger $k$-bins. While large ($1$-$30$ \%), several of the relative signals shown in Fig.~\ref{fig:halopowerspecsWmarks} are approximately constant on these scales; in this case, the effect can be absorbed into a redefinition of the halo bias $b$. Breaking this degeneracy requires additional measurements, for example through Redshift-Space Distortions (RSD). A more comprehensive analysis accounting for this is interesting but beyond the scope of the present work. The marked spectra seem promising because they give moderate to large enhancements of the signal, but the larger jackknife errors indicate that the effect of shot noise is amplified, requiring either larger tracer catalogues, or a better treatment of the nonlinear shot noise.
 
    \section{Summary and conclusion}

    The case of an inhomogeneous and structure-induced phase transition (SIPT), where
    $z_{\rm{SSB}}\gg z_{*}$, presents us with an interesting model case to confront
    with observational data, where environment-dependent effects are in the front seat. Although
    the symmetron model was studied in particular here, our results should be
    kept in mind when developing other late-time phase transition models (with e.g.\ non-perturbative
    potentials or conformal factors).
    Our main findings can be summarised as follows.
    \begin{itemize}
        \item[] \textbf{\small Model} 
        \item SIPT-type models may be applied to address the coincidence problem of dark energy by coupling the phase transition to nonlinear structure formation. But achieving energy densities relevant for dark energy requires non-perturbative conformal factors and/or potentials, as discussed in Sections \ref{S:introduction}-\ref{S:method}.
        \item The SIPT-type symmetron can generate very structured and small-scale defect networks that pin to overdensities and do not enter the scaling regime. They can therefore maintain a high defect density with cosmic time. Their late formation and relatively low energy scale avoids issues with overclosing of the Universe.
        \item The defect densities in SIPT-type models depend on the densities of initially isolated voids that are sufficiently underdense at the time of its phase transition. This is set by both the phase transition energy $\rho_*$ (or $z_*$) and the Compton wavelength $L_C$, see e.g. Figs.~\ref{fig:overview}-\ref{fig:defect_renderings}. The defect density is insensitive to the causal horizon and is therefore different from the usual phase transition models that make use of the Kibble mechanism.
        \item $\phi\in\mathbb{R}$ tends to have better screening properties than $\phi\in\mathbb{C}$, due to the larger volume coverage of the wall defects compared to the string defects, and steeper field profiles when approaching overdensities (see Fig.~\ref{fig:idealString} for defect profiles).
        \item Large defect densities can be achieved and applied to enter a regime where topological screening is more important, presumably to a much greater extent than what was explored here.
        \item[] \textbf{\small Phenomenology} 
        \item We have observed a migration of defects to locations with greater ambient densities over time (see Fig.~\ref{fig:stringenvhistogram}).
        \item The strong coupling/strong
    screening regime tends to have large effects on the velocity dispersion of
    cold dark matter particles (Fig.~\ref{fig:CDMvelocities}) and smaller but still large effects on the halo velocity dispersion (Fig.~\ref{fig:HalosVelocities})
    as the halos tend to form in the higher density, screened regions.
        \item  The  halo formation
    is mostly enhanced due to the fifth force enhancing clustering and increasing the matter infall into overdense
    regions. However, in some cases, we are seeing the formation of halos
    suppressed, presumably due to the strong acceleration of particles in
    underdense regions, which is more pronounced for $\phi\in\mathbb{C}$ that has
    less steep screening and defect profiles, see e.g.\ Fig.~\ref{fig:idealString}. However, this interpretation requires a more careful analysis to verify, see the discussion in Section \ref{SS:densityDistribution}.
        \item Underdense regions where the
    phase transition is occurring are dramatically more empty than they would be
    in $\Lambda$CDM (Fig.~\ref{fig:densityAll}), which is visible in the low-density tail of the PDF, Fig.~\ref{fig:hist1pdensity}.
        \item Different models can affect different probes very differently. For example model $3_{\mathbb{R}}$ where the global dark matter power spectrum is
    only changed with respect to $\Lambda$CDM at a $\sim 4 \%$ level (Fig.~\ref{fig:pk_matter}), the halo mass function shows $40\%$ effects for small masses. We see 20-25\% effects on the halo spectra at $z=0$, while
    it is sub-percent at $z=0.3$.
        \item A flat enhancement of the halo power spectrum can be absorbed into the linear halo bias parameter, requiring additional probes, such as RSD, to disentangle the degeneracy. 
        \item[] \textbf{\small Environment-dependent probes} 
        \item As discussed in Subsection \ref{SS:probes},
    effects that appear exclusively at low redshift are observationally more challenging due
    to the smaller survey volumes. This shows up as large sample variance in both of our environment-dependent probes.
        \item Assuming that the effects we found generalise to the cases of galaxy spectra and the PDF estimators, it is likely that the full set of models considered here can be constrained by currently available data.
        \item Less extreme choices of the matter coupling $\beta$ will be consistent with the observed galaxy power spectra and other standard probes while giving strong effects in underdense regions, motivating the continued development of environment-dependent probes such as one-point PDFs, marked statistics and analyses based on voids.
        \item SIPT-type models like the symmetron make an interesting test case for the development of such probes, and may be used as a proxy for other models inducing strong environment-dependent effects in the late-time Universe.
    \end{itemize}

   In conclusion, we have shown that SIPT-type dark sector phase transitions can couple modifications of $\Lambda$CDM cosmology to nonlinear structure formation, generating dense and persistent string or wall defect networks that evade standard Kibble scaling. The phase transition induces strong environment-dependent effects, which we investigate using the matter PDF and marked halo statistics. These models can leave only modest imprints on traditional large-scale observables while producing dramatic modifications in underdense regions, halo properties, and velocity statistics at late times. As a result, the inclusion of environment-dependent probes is expected to allow strong constraints on even more subtle parameter choices (such as smaller coupling $\beta$). In particular the detection of the low-density tail of the PDFs emerges as a key observational target. Observables such as matter density PDFs, marked statistics, and void-based analyses are therefore essential. SIPT-type models provide a compelling and well-controlled framework for developing and testing such probes, with broader relevance for any late-time physics that couples the dark sector to underdense environments.

    \small{\textit{\paragraph{Acknowledgements:} \O{}.C.\ thanks Hans Winther for useful discussions. This work was co-funded by the European Union and supported by the Czech Ministry of Education, Youth and Sports (Project No. FORTE – CZ.02.01.01/00/22\_008/0004632). J.A.\ acknowledges funding by the Swiss National Science Foundation and the Dr.\ Tomalla Foundation for Gravity Research. M.K.\ acknowledges funding by the Swiss National Science Foundation. This work was supported by a grant from the Swiss National Supercomputing Centre (CSCS) under project ID sm97 on Alps, and by the Ministry of Education, Youth and Sports of the Czech Republic through the e-INFRA CZ (\text{PiD}:FTA-25-52).}}

    \bibliographystyle{JHEP}
    \bibliography{inspire,not_in_inspire}

    \appendix

    \section{String profile}
    \label{A:stringProfile}

    In order to validate the implementation of the model in the code, we start
    by solving a simple string on a homogeneous background. By assuming an axisymmetrical
    system, following \cite{durrer_topological_1999}, also covered in \cite{Nezhadsafavi:2025pzg,Hindmarsh:1994re},
    one can semi-analytically solve the scalar profile of a string. One finds the
    quasistatic equation of motion, where the Laplacian is put in polar (and comoving)
    coordinates
    \begin{equation}
        \partial_{r}^{2} \phi + \frac{1}{r}\partial_{r} \phi + \frac{1}{r^{2}}\partial
        _{\theta}^{2} \phi = a^{2} V_{\phi} .
    \end{equation}
    If we make the Nielsen-Olesen ansatz \cite{Nielsen:1973cs} that the
    only azimuthal dependence is for the phase, which does not have radial dependence,
    then $\phi= v f(r) e^{i n \varphi}$, where $\varphi$ is the phase, $n$ is
    the winding number, $v\equiv \frac{\mu}{\sqrt{\lambda}}$ is the vacuum expectation
    value, while $f$ is the radial field profile. Defining the radial coordinate
    $s=r v$ gives the equation for the profile
    \begin{equation}
        f_{,ss}+ \frac{1}{s}f_{,s}- \frac{n^{2}}{v^{2}}f - a^{2} \lambda f \left(
        f^{2} -\left[1-\bar\rho_{m}/\rho_{*}\right] \right) = 0 .
    \end{equation}
    We can solve this as an initial boundary value problem with $f(r=\infty),f_{,s}
    (r=\infty) = v,0$ and $f(r=0),f_{,s}(r=0)=0,0$. We put the winding number $n=1$. 
    \begin{figure}[tb]
        \centering
        \includegraphics[width=\linewidth]{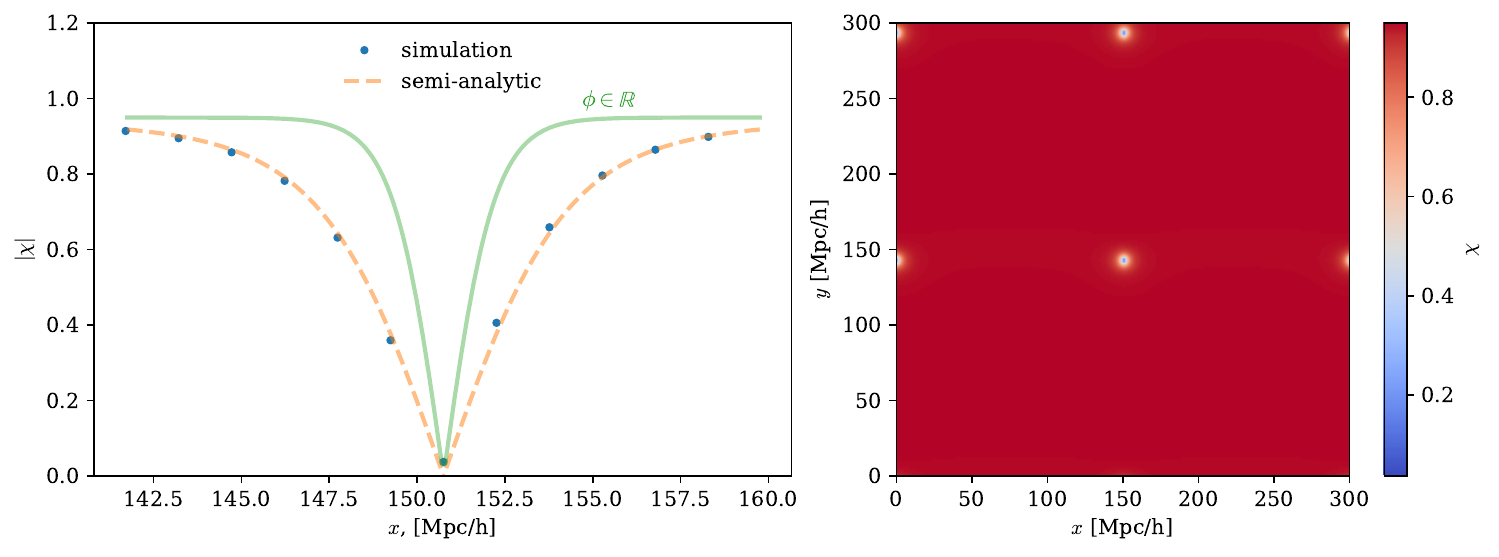}
        \caption{Quasistatic field $|\chi|=|\phi|/v_{0}$ after relaxing it using
        Gauss-Seidel relaxation. Left panel: field profile of centre-most string
        defect, orthogonal to the defect axis. We compare with the semi-analytic
        profile found from the ODE derived in \cite{durrer_topological_1999}. Green line shows the analytic profile of the real-valued field $\phi\in\mathbb{R}$.
        Right panel: grid of strings formed after the relaxation.}
        \label{fig:idealString}
    \end{figure}
    For the symmetron parameters $(L_C,\;z_*,\;\beta) = (1\;h^{-1}\mathrm{Mpc},\;2,\;8)$, we find the profile shown in the left panel of Fig.~\ref{fig:idealString}. For the simulation, we initialise the scalar field in four quadrants at $z=0.4$ at its background minimum $|\phi|=v_0\sqrt{1-\rho_\mathrm m (z)/\rho_*}$. We use a homogeneous denisty field $\rho_\mathrm m (z)=\Omega_\mathrm m\;\rho_{\mathrm c, 0}/a^3$. The four quadrants are initialised at different phase angles $\varphi=\pi/2,\;\pi,\;0,\;-\pi/2$, from upper right quadrant and clockwise. Then Gauss-Seidel relaxation (see \cite{Christiansen:2023tfy}) is iterated until a tolerance $|\chi_{i+1}-\chi|_{\mathrm {max}}<10^{-8}$ is reached. This is done in a boxsize of $B=300\;h^{-1}\mathrm{Mpc}$, and a spatial resolution $\mathrm d x=1.5\,h^{-1}\mathrm{Mpc}$. This is less than our spatial resolution criterion $\mathrm d x/L_C<1$ that we keep in our dynamical simulations. Figure~\ref{fig:idealString} shows string defects forming after relaxation at a regular grid at the intersections of the four quadrants (with periodic boundary conditions). The string profile is seen to match well with the analytical expectation (left panel).

    \section{Convergence}
    \label{A:convergence}

    We present here a convergence analysis to indicate that the general results
    we are seeing come from the physical model, which is being solved with appropriate
    temporal and spatial resolution. We are, however, still seeing some errors
    on selected quantities which we discuss the significance and cause of.

    \begin{figure}[tb]
        \centering
        \includegraphics[width=0.7\linewidth]{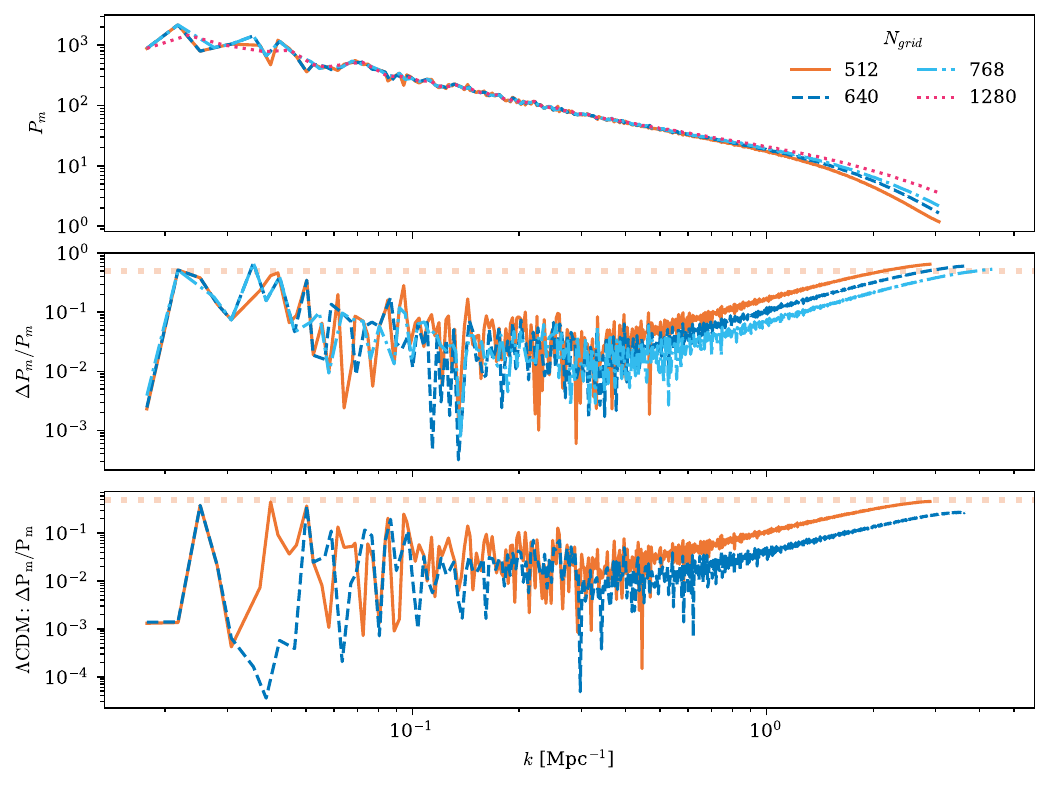}
        \caption{ Convergence of matter power spectrum with increasing resolution.
        Upper pane shows the matter power spectra for the symmetron model
        $1_{\mathbb{C}}$, with the middle pane given the relative difference with
        respect to the $N=1280^{3}$ simulation. The lower pane shows the
        relative difference in the analogue case but for $\Lambda$CDM simulation. }
        \label{fig:pkm_conv}
    \end{figure}

    In Fig.~\ref{fig:pkm_conv}, we show the spatial convergence in the matter power spectrum
    for model 1 (top and middle panels) and $\Lambda$CDM (bottom panel). We see convergence
    in both cases, though faster for the $\Lambda$CDM case. For example, the error
    on $k=2\,h\,\mathrm{Mpc}^{-1}$ for $N=512^{3}$ increases by a factor $\sim 1.
    4$ for the symmetron case, with respect to $\Lambda$CDM. Since the realisation is not the same at different spatial resolutions, we are seeing the effect of sample variance at small $k$.

    \begin{figure}[tb]
        \centering
        \includegraphics[width=0.7\linewidth]{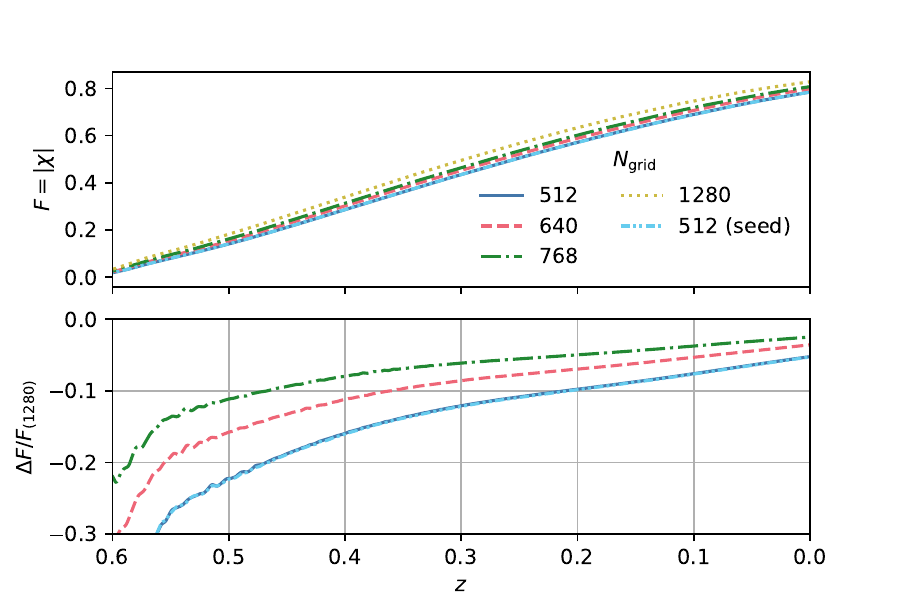}
        \caption{ Convergence of box averaged norm of the scalar field $|\chi|=\sqrt{\chi^{\dagger}\chi}$
        for the model $1_{\mathbb{C}}$ at different spatial resolutions. `seed' indicates that the simulation is run with a different seed number. }
        \label{fig:conv_achiav}
    \end{figure}

    Figure~\ref{fig:conv_achiav} shows the time evolution of the volume averaged
    norm of the scalar field $\sqrt{\chi^{\dagger} \chi}$. The exact time of start
    of the phase transition is very sensitive to the resolution, allowing a
    slightly earlier phase transition in the more highly resolved case, causing
    a slight horizontal displacement, approaching a $5\%$ error towards the
    present time, which improves with higher resolution. We note that sample
    variance is not important here, as the different seed number has no effect.
    Although the relative error is decreasing, we see an almost constant absolute error after $z\sim 0.5$, owing to the time displacement of the different resolution simulations.

    \begin{figure}[tbp]
        \centering
        \begin{subfigure}
            [b]{0.5\textwidth}
            \centering
            \includegraphics[width=\textwidth]{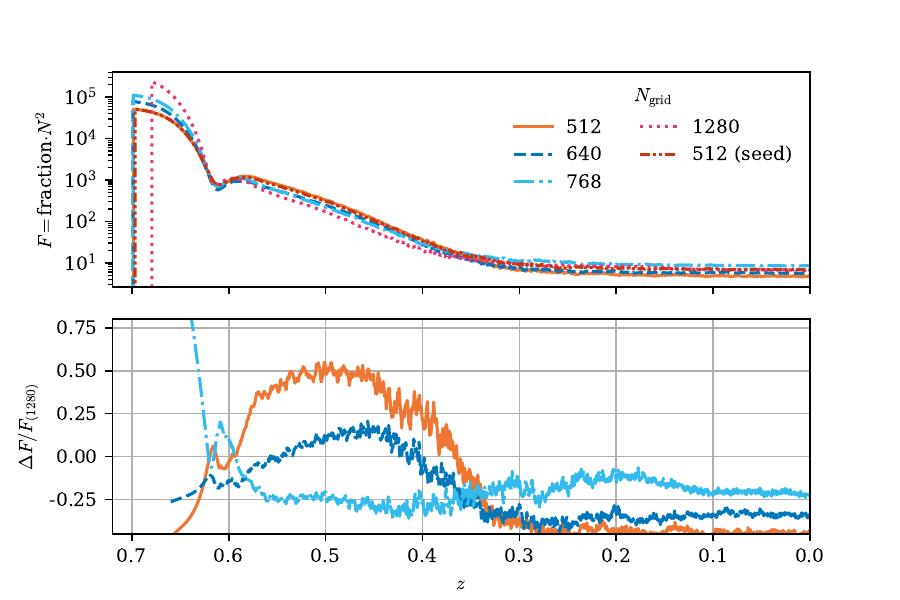}
        \end{subfigure}
        \hspace{-0.05\textwidth}
        \begin{subfigure}
            [b]{0.5\textwidth}
            \centering
            \includegraphics[width=\textwidth]{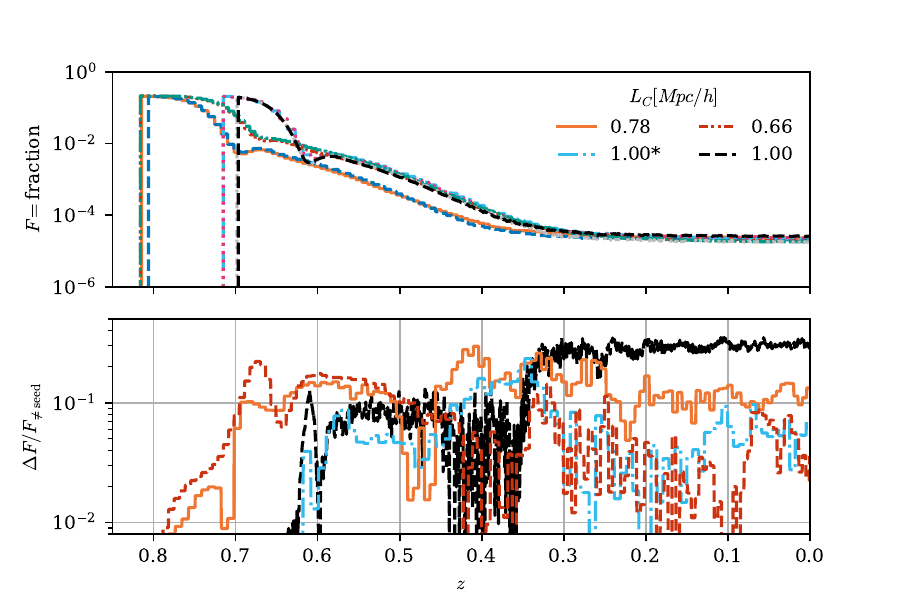}
        \end{subfigure}
        \caption{Fraction of volume of simulation occupied by strings. Left panel:
        The case of model $1_{\mathbb{C}}$, with $z_{*},\; \beta,\; L_{\mathrm{C}}=0.
        91,\; 16,\; 1\,h^{-1}\,\mathrm{Mpc}$. Box fraction is multiplied
        $N_{\rm{grid}}^{2}$ for correct scaling of 1-dimensional objects. Right panel: Comparison for
        pairwise simulations with different seed numbers, for differently chosen
        Compton wavelengths, indicating the effect of sample variance. Asterisk on
        green graph label indicates that the pair is done in a larger box $B=750\,
        h^{-1}\,\mathrm{Mpc}$.}
        \label{fig:conv_sfracav}
    \end{figure}

    Finally, in Fig.~\ref{fig:conv_sfracav}, we check for errors on the observed
    string density in the box. The relevant quantity that would be expected to
    be consistent in fully resolved simulations is the string energy density, $\rho
    _{\mathrm{string}}$, which relates to the fraction of lattice points
    occupied by strings, $F$, as
    \begin{equation}
        \rho_{\mathrm{string}}= F \mu / \mathrm{d x}^{2},
    \end{equation}
    where $\rm d x$ is the spatial resolution of the simulation and
    $\mu\equiv \int_0^{2\pi} \mathrm d \theta\int_{-\infty}^{\infty} r\mathrm{d}r \enspace V(\phi)$
    is the string tension. In particular on this convergence check, we notice a
    significant impact of changing the seed number of the simulation; see the left
    panels. This indicates that the error might be dominated by sample variance,
    since the box volume is only forming a few strings in model $1_{\mathbb{C}}$,
    which we see to be the case in Fig.~\ref{fig:defect_renderings}. Since we do
    not have concordance between phase realisation at the different spatial
    resolutions, in effect, every resolution level has a slightly different
    number of strings, and different number of string annihilations and interactions.
    This indicates that model $1_{\mathbb{C}}$ errors might improve on
    simulating a larger volume or different parameters with smaller string
    spacings. However, the error we see is still less than $\sim 25\%$, apart from
    in the very early stages of the phase transition. In the right panels of Fig.~\ref{fig:conv_sfracav},
    we see that decreasing the Compton wavelength, which we know reduces the string
    separation (see Fig.~\ref{fig:defect_renderings}), or increasing the volume while
    keeping resolution constant, reduces the effect of sample variance on from
    $\sim 30\%$ to $5$-$10\%$. The height of the initial plateau immediately after the phase transition varies with spatial resolution, as shown in the upper left panel of Fig.~\ref{fig:conv_sfracav}. At early times, defect formation is characterised by small field amplitudes and spurious phase discontinuities identified by the defect finder, which later merge into coherent structures. These initial plateaus can be made to overlap by removing the $N^2$ factor, which assumes that physical strings are one-dimensional objects and therefore scale as $F\sim N^{-2}$. This indicates that the initial defects are instead formed as three-dimensional collapsing regions in underdensities, which subsequently evolve into extended one-dimensional strings. These three-dimensional collapse regions are therefore resolved with a different scaling behaviour.

    In conclusion, spectra and background quantities are found to have little
    effect from sample variance and are both converging with spatial resolution.
    The initially large but decreasing error of background quantities are likely caused by sensitivity of the time of the start of the phase transition on the spatial resolution, as underdensities are available earlier in the more finely resolved
    simulations. The string energy density is resolved, assumedly up until
    sample variance, which is found to be decreasing with more ergodic samples.
    For quantities that may be sensitive to string density, such as gravitational
    wave amplitudes (when dominated by defect emission), there might be a
    corresponding error of $5$-$30\%$, depending on the sample choice. This
    sensitivity is, as expected, higher than in the case of domain walls \citep{Christiansen:2024vqv,Christiansen:2024uyr},
    which cover more of the simulation volume. The error will improve by increasing
    the dynamical range of the simulations.

    \section{Performance}
    \label{A:performance}

    The main simulations were run on the Eiger supercomputer at the Swiss National
    Supercomputing Centre (CSCS) using 50 nodes of $2\times$AMD 7742 chips,
    giving a total of 6400 cores. The simulation domain was decomposed into $80\times
    80$ rods. We observe the runtime costs as
    \begin{align}
        \phi\in \mathbb{C}: & \enspace 30.6\times 10^{3}\times \left(0.41+0.59\times\frac{0.07}{C^{(\phi)}}\right)\left(\frac{N}{1280}\right)^{4}\times\left(\frac{1}{C^{(\mathrm{cdm})}}\right)\,\mathrm{CPU\text{-}h}, \\
        \phi\in \mathbb{R}: & \enspace 32.2\times 10^{3}\times \left(0.39+0.61\times\frac{0.07}{C^{(\phi)}}\right)\left(\frac{N}{1280}\right)^{4}\times\left(\frac{1}{C^{(\mathrm{cdm})}}\right)\,\mathrm{CPU\text{-}h},
    \end{align}
    plus the output cost, which was an additional $10\%$ for the output produced
    in this case, of which the main expense was the I/O related to animations and
    snapshots. We have conservatively assumed a $N^{4}$ scaling, as the run time
    is dominated by lattice updates $\propto N^{3}$ and we assume a constant
    Courant factor – $C^{(\mathrm{cdm})}$ and $C^{(\phi)}$ for CDM and the
    scalar field, respectively – which means that the number of time steps is $\propto
    N$. We see a more or less similar performance between the
    $\phi\in\mathbb{C}, \mathbb{R}$ cases. The case of $\Lambda$CDM can be
    recovered for $C^{(\phi)}\rightarrow \infty$. $90\%$ of the scalar field sector's
    runtime is coming from the field equations solver, which makes it the best
    candidate for optimisations. Possible improvements might be offered by applying
    implicit solvers instead of the $4$th order Runge-Kutta solver currently
    used, although the relevant physical scales should be considered carefully,
    see \cite{Christiansen:2024vqv}. This sort of physics problem,
    requiring a large number of timesteps for the rapidly oscillating scalar, will
    in particular benefit from working on GPUs, which is the subject of a future
    work.
\end{document}